\documentclass[lettersize,journal]{IEEEtran}
\usepackage{amsmath,amsfonts}
\usepackage{algorithmic}
\usepackage{algorithm}
\usepackage{array}
\usepackage[caption=false,font=normalsize,labelfont=sf,textfont=sf]{subfig}
\usepackage{textcomp}
\usepackage{stfloats}
\usepackage{url}
\usepackage{verbatim}
\usepackage{graphicx}
\usepackage{cite}
\usepackage{standalone}
\usepackage{tikz}
\usepackage{bm}
 \usepackage{booktabs} 
\usetikzlibrary{positioning}
\usepackage{multirow}

\usepackage{pifont}% http://ctan.org/pkg/pifont
\newcommand{\cmark}{\ding{51}}%
\newcommand{\xmark}{\ding{55}}%

\everymath=\expandafter{\the\everymath\displaystyle}

\definecolor{input}{RGB}{210, 255, 210}
\definecolor{context}{RGB}{255, 160, 0} 
\definecolor{current}{RGB}{100, 150, 255}
\definecolor{unknownFill}{RGB}{245, 245, 245}
\definecolor{unknownBorder}{RGB}{180, 180, 180}

\newcommand{\drawNTapFilter}[2]{
\begin{tikzpicture}[scale=#1]
    \def\cols{9}
    \def\rows{6}
    \def\currentIdx{40} 

    \ifnum#2=3 \def\taps{30, 31, 39} \fi
    \ifnum#2=5 \def\taps{22, 30, 31, 38, 39} \fi
    \ifnum#2=8 \def\taps{13, 22, 30, 31, 32, 37, 38, 39} \fi
    \ifnum#2=16 \def\taps{13, 14, 20, 21, 22, 23, 24, 28, 29, 30, 31, 32, 33, 37, 38, 39} \fi

    \foreach \i in {0,...,53} {
        \pgfmathsetmacro{\ix}{int(mod(\i,\cols))}
        \pgfmathsetmacro{\iy}{int((\rows - 1) - floor(\i/\cols))}
        \ifnum\i=\currentIdx
            \fill[current] (\ix,\iy) rectangle ++(1,1);
            \draw[black, thick] (\ix,\iy) rectangle ++(1,1);
        \else\ifnum\i<\currentIdx
            \fill[input] (\ix,\iy) rectangle ++(1,1);
            \draw[gray!40] (\ix,\iy) rectangle ++(1,1);
        \else
            \fill[unknownFill] (\ix,\iy) rectangle ++(1,1);
            \draw[unknownBorder] (\ix,\iy) rectangle ++(1,1);
        \fi\fi
    }

    \foreach \t in \taps {
        \pgfmathsetmacro{\tx}{int(mod(\t,\cols))}
        \pgfmathsetmacro{\ty}{int((\rows - 1) - floor(\t/\cols))}
        \draw[black, line width=1.1pt] (\tx,\ty) rectangle ++(1,1);
        \fill[context, fill opacity=0.4] (\tx,\ty) rectangle ++(1,1);
        \fill[context] (\tx+0.05,\ty+0.05) rectangle ++(0.9,0.9);
    }

    \pgfmathsetmacro{\cx}{int(mod(\currentIdx,\cols))}
    \pgfmathsetmacro{\cy}{int((\rows - 1) - floor(\currentIdx/\cols))}
\end{tikzpicture}
}

\begin{document}

\title{LANCE: Locally Adaptive Neural Context Estimation for Overfitted Image Compression}

\author{Martin Benjak and Jörn Ostermann,~\IEEEmembership{Fellow,~IEEE}
        % <-this % stops a space
\thanks{Martin Benjak and Jörn Ostermann are with the Institut für Informationsverarbeitung, Leibniz Universität Hannover, 30167 Hannover, Germany (e-mail:
benjak@tnt.uni-hannover.de).}% <-this % stops a space
%\thanks{Manuscript received April 19, 2021; revised August 16, 2021.}
}

\markboth{}%
{Shell \MakeLowercase{\textit{et al.}}: A Sample Article Using IEEEtran.cls for IEEE Journals}

\IEEEpubid{\begin{tabular}[t]{@{}c@{}}Copyright \copyright~2026 IEEE. Personal use of this material is permitted.\\However, permission to use this material for any other purposes must be obtained from the IEEE by sending an email to pubs-permissions@ieee.org.\end{tabular}}

\maketitle

\begin{abstract}
This paper introduces Locally Adaptive Neural Context Estimation (LANCE), a novel extension for overfitted image compression (OIC) frameworks like Cool-Chic. While traditional OIC methods rely on lightweight autoregressive networks with globally signaled parameters, they struggle with non-stationary image statistics. LANCE addresses this by incorporating a forward-signaled spatial hyperprior that enables regional adaptation of the entropy model. To minimize overhead, we employ a predictive coding scheme that combines a static Median Edge Detector (MED) with a lightweight learned context model.

Experiments demonstrate that LANCE achieves BD-rate reductions of 1.40\% on the Kodak dataset and 1.97\% on CLIC 2020 over Cool-Chic 4.0 at the high end of our decoder complexity range of 606-1483 MAC/pixel. At the low end of the complexity range, we outperform Cool-Chic 4.0 by 2.41\% and 2.99\% on Kodak and CLIC, respectively. Qualitative analysis reveals that the learned spatial hyperprior effectively segments image regions into areas of similar image statistics, providing an automated, content-aware adaptation layer.
\end{abstract}


\section{Introduction}
\label{introduction}
The landscape of image compression has shifted dramatically in recent years. Traditionally, conventional codecs like JPEG \cite{pennebaker1992jpeg} and HEVC  \cite{hevc_ohm} have relied on handcrafted coding tools and sophisticated adaptive entropy coding methods such as CABAC \cite{cabac} to adapt to varying image statistics. More recently, learned codecs have leveraged deep neural networks to optimize the entire compression pipeline end-to-end. A specialized subset of these are overfitted codecs \cite{dupont2021coin, ladune2023cool, leguay2024cool}, which optimize and signal network parameters for each specific image.

A central element of each image codec is the entropy coder. Entropy coding rests upon Shannon’s Source Coding Theorem, which establishes that the average length of a code cannot be smaller than the entropy $\mathrm{H}(X)$ of the source $X$ with symbols $x_i \in X$. When the source symbols exhibit dependencies and $C$ is a context representing previously coded symbols, the conditional entropy $\mathrm{H}(X|C)$ is often significantly lower than the entropy $\mathrm{H}(X)$. A static coder that uses a fixed probability distribution is suboptimal in these cases. Adaptive entropy coders have the ability to modify their probability model $P(X)$ in response to the observed characteristics of the data sequence being processed. This adaptability can generally be implemented in two ways: forward and backward. Forward adaptation involves the encoder analyzing a block of data, determining the optimal model parameters, and explicitly signaling those parameters to the decoder as side information. Backward adaptation allows the decoder to infer the model parameters directly from the previously decoded symbols. Backward adaptation thus ensures that both the encoder and decoder remain synchronized without the need for additional overhead bits, provided they follow the same deterministic rules for updating their internal states.

\IEEEpubidadjcol

JPEG \cite{pennebaker1992jpeg} uses a purely forward adaptive entropy model by allowing the signaling of Huffman tables used by its entropy coder in the bitstream. Modern video coding standards such as AVC, HEVC \cite{hevc_ohm} and VVC \cite{bross2021developments_VVC} use CABAC \cite{cabac} as their entropy coder. CABAC is backward adaptive by updating discrete context models based on the statistics of previously decoded symbols.

\begin{figure}[!t]
    \centering
    \input{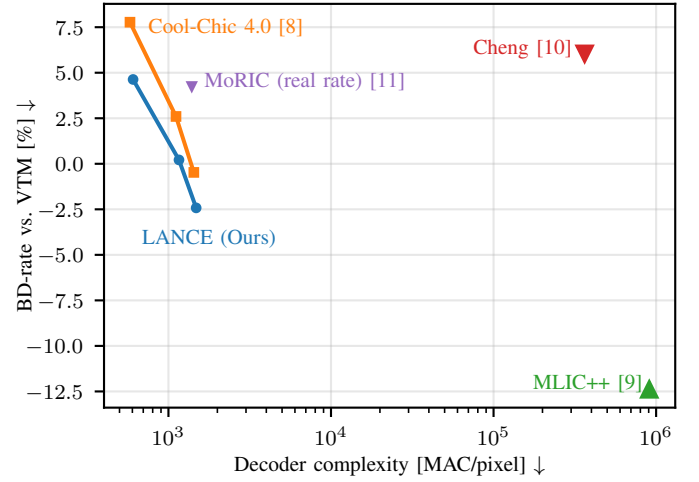}
    \caption{Rate savings in terms of BD-rate versus VVC (VTM 23.14) over decoder complexity evaluated on the CLIC 2020 professional
validation set \cite{clic2020}. Negative results: less rate is required to get the same quality as VVC.}
\label{fig:complexity_clic}
  \end{figure}

Learned image compression (LIC) has developed by progressing through three distinct stages of adaptivity: \textit{static}, \textit{forward adaptive}, and \textit{joint forward and backward adaptive}. The first fully end-to-end trained LIC scheme by Ballé et al. \cite{balle2016end} utilized a \textit{static} model, which relied on a fixed prior and assumed that the latent representations were statistically independent, limiting the coder's ability to exploit spatial correlations. This was followed by the introduction of \textit{forward adaptation} through the scale hyperprior \cite{balle2018variational}. This method extracts side information to predict the distribution parameters of the latent, though it necessitates signaling this side information to the decoder. Modern state-of-the-art architectures now utilize \textit{joint forward and backward adaptation}, where the hyperprior provides high-level side information (forward) while an autoregressive context model utilizes the decoded values of causal neighbors to provide local information (backward) \cite{minnen2018joint}. In these joint models, the distribution parameters are conditioned on both the signaled hyperprior and the causal context, allowing for a probabilistic model with significantly higher accuracy, that effectively minimizes the discrepancy between the estimated distribution and the actual image statistics.

Overfitted image compression (OIC) represents a further specialization of these adaptive principles, where the entire coding framework is optimized for the statistics of a single image instance rather than a broad dataset. Modern frameworks like Cool-Chic \cite{ladune2023cool} and C3 \cite{kim2024c3} differ from conventional LIC by skipping the general-purpose analysis transform. Instead, they derive image-specific representations through a gradient-based optimization process that jointly overfits three components to a single image: hierarchical latent grids, a lightweight synthesis transform, and an autoregressive context model. In Cool-Chic, the majority of the image information is stored in a multi-layer pyramid of latent grids at different spatial resolutions. These latents are upsampled and synthesized by the overfitted synthesis transform to recover RGB values, while their probability distributions are estimated by the image-specific autoregressive context model. This context model predicts the mean and variance parameters of a probability density function for each latent value based on a causal context providing a highly granular form of backward adaptation. The weights of the context model convey information about the global image statistics, similar to the hyperprior found in conventional LIC. By transmitting these quantized weights, the system provides a form of forward adaptation that parametrizes the autoregressive context model to global image statistics.

However, a challenge remains: images are rarely homogeneous, with statistical properties varying drastically within a single frame. Consider a screen capture containing both sharp text and natural textures, or a complex photo montage. Images should thus be modeled as a non-stationary random field indicating the need to adapt the functionality of the context model to local changes in image statistics. 

The existing adaptation mechanism by signaling image specific weights for the context model can only adapt the context model to global image statistics. To intrinsically adapt to complex, spatially non-stationary statistics, an autoregressive context model would require a large input context and a very expressive network architecture with a high parameter count. However, OIC frameworks generally utilize relatively shallow and lightweight auto-regressive networks to keep the signaling cost of the network weights low. These shallow networks are unable to adapt to local changes in the image statistics.

To address this limitation, we introduce Locally Adaptive Neural Context Estimation (LANCE) which offers another layer of adaptability to the autoregressive context model used in OIC by introducing a forward signaled spatial hyperprior which is an adaptation of the concept of the hyperprior from LIC \cite{balle2018variational} to the requirements of OIC. The spatial hyperprior models local changes in non-stationary image statistics and adapts the autoregressive context model accordingly. We assume that the parameters of the spatial hyperprior exhibit strong spatial correlations. Thus, we utilize predictive coding to efficiently compress and signal the spatial hyperprior parameters.

To sum up, this work has the following contributions:
\begin{itemize}
    \item We introduce a learned spatial hyperprior that provides forward signaled side-information to the context model within a learned overfitted image codec to adapt to local changes in image statistics.
    \item We introduce an additional hybrid autoregressive context model to efficiently compress and signal the parameters of the spatial hyperprior, assuming that its parameters have strong spatial correlations.
    %\item LANCE achieves a balance between forward and backward adaptation, allowing for more expressive entropy estimation without substantial increases in complexity.
    \item We allow parallel decoding, by not introducing any inter latent layer dependencies. 
    \item We demonstrate that LANCE improves the compression performance to decoding complexity trade-off of overfitted image codecs. Specifically, the compression performance is improved by $2.99\%$ at a low decoding complexity of $606 \text{ MAC}/\text{pixel}$, and by $1.97\%$ at a high decoding complexity of $1483 \text{ MAC}/\text{pixel}$ in terms of BD-rate compared to Cool-Chic 4.0 on the CLIC 2020 professional validation set \cite{clic2020}.
\end{itemize}

The remainder of this paper is organized as follows:
Section \ref{related_work} reviews the recent progress in learned image compression with a focus on overfitted image compression. Section \ref{method} presents our LANCE Method. Section \ref{evaluation}
addresses the ablation experiments, rate-distortion comparison, and complexity analyses. Finally, we provide concluding remarks in Section \ref{conclusion}.

\section{Related Work}
\label{related_work}
Most lossy image and video codecs since H.261 and JPEG \cite{pennebaker1992jpeg} have been based on the transform coding scheme, in which signals are encoded by quantization and entropy coding in the transform domain rather than the pixel domain. The aforementioned codecs, as well as more modern ones such as JPEG XL \cite{alakuijala2019jpeg}, HEVC \cite{hevc_ohm}, and VVC \cite{bross2021developments_VVC}, use linear transformations like the DCT or DST. While LIC schemes also follow the transform coding paradigm, they replace modules such as the transform and entropy model with nonlinear neural networks. These networks are trained using gradient descent and a rate-distortion loss function on large, varied image datasets with the goal of creating a single codec that performs well across many different images. In 2025, the first LIC standard JPEG-AI \cite{alshina2024jpeg} was published which outperforms VVC in terms of perceptual quality while operating at a low decoder complexity in the range of 8 to 215 MAC/pixel. In contrast, OIC methods overfit codec parameters to individual image instances offering a good rate-distortion performance with lightweight decoders below 2000 MAC/pixel compared to most LIC methods that require up to 1 million MAC/pixel. The following subchapters describe the development of LIC and OIC and set our method LANCE in context with other state-of-the-art methods.   
\subsection{Learned Image Compression}
The first practical end-to-end learned image compression scheme was introduced by Ballé et al \cite{balle2016end}, which utilized linear transforms and generalized divisive normalization \cite{balle2015density} as nonlinearity to optimize the rate-distortion trade-off. To improve the entropy encoding of the latent space, Ballé subsequently introduced the hyperprior \cite{balle2018variational}, a second hierarchical level of latent variables that provides side information to predict the distribution parameters of the primary latents. Minnen et al. \cite{minnen2018joint} extended this idea with an autoregressive context model, which utilizes the decoded values of causal neighbors to further refine the probabilistic estimation of each symbol. This combination of nonlinear transforms with joint backward and forward adaptation using an autoregressive context model and a hyperprior is the foundational architecture for most modern learned image compression frameworks.

Cheng et al. \cite{cheng2020learned} extended this base architecture with discretized Gaussian mixture likelihoods and attention modules to parameterize complex latent distributions, achieving performance comparable to VVC \cite{bross2021developments_VVC}. Minnen\cite{minnenChannelwiseAutoregressiveEntropy2020} et al. introduced a channel-wise autoregressive context model which was extended by He et al. to the spatial-channel context model of ELIC \cite{he2022elic}. Cao et al. \cite{cao2025generative} use denoising diffusion implicit models (DDIMs) \cite{song2020denoising} to map the hyperprior and causal context to parameters of a probability distribution. Recently, further gains were achieved by fusing diverse context information within the entropy model \cite{qian2022entroformer, jiang2025mlic, li2025learned}:
Qian et al. \cite{qian2022entroformer} explored long-range spatial context modeling with transformer architectures; MLIC++ \cite{jiang2025mlic} and HPCM \cite{li2025learned} exploit both local and global long-range correlations using a multi-reference and a progressive multi-scale context fusion mechanism, respectively. These developments demonstrate that sophisticated entropy parameter modeling is crucial for improving the performance of image coding schemes. However, these methods focus on more-and-more diverse backward causal context information to adapt the entropy model while the forward information is still signaled using a simple hyperprior.

Kim et al. \cite{kim2022joint} introduce a multi-head attention-based global hyperprior additional to the local hyperprior. By this, they can effectively model different types of dependencies between latent elements. \cite{kim2024diversify} extends this idea even further by introducing a third regional hyperprior.
These methods are conceptually similar to our LANCE method, where the weights of the entropy model are signaled analogous to a global hyperprior and the spatial hyperprior captures local variations of the image statistics. All hyperpriors in \cite{kim2022joint} and \cite{kim2024diversify} are encoded using non-linear transform coding and a static non-parametric fully factorized entropy model as in \cite{balle2016end}. This is not suitable for OIC, since a non-linear transform has to be signaled. Therefore, we take a different route by encoding our spatial hyperprior using predictive coding with an autoregressive context model that takes dependencies of neighboring elements into account.

\subsection{Overfitted Image Compression}
Implicit Neural Representations (INR) describe methods that represent data using neural networks. They were originally introduced by Stanley \cite{stanley2007compositional} for DNA data. Later, the idea was adapted to represent other forms of high-resolution data such as shape representations \cite{park2019deepsdf} and neural radiance fields \cite{mildenhall2021nerf}. Dupont et al. extended this idea into the image compression domain by introducing COIN \cite{dupont2021coin}. It uses a lightweight Multi-Layer Perceptron (MLP) with sine activations \cite{sitzmann2020implicit} to map the coordinates of a pixel to its RGB values. To encode an image, the MLP is overfitted to the pixels of a single image and then quantized and transmitted. Thus, the image information is encoded in the MLP itself. COIN++ \cite{dupont2022coinpp} meta-learns a base-network that is shared between all image instances and overfits a latent modulation that is quantized and signaled to the decoder-side. Compared to LIC methods, COIN and COIN++ have a low decoder complexity. However, their rate-distortion performance is below JPEG and therefore not on par with computationally more complex LIC. Stümper et al. \cite{strumpler2022implicit} proposed the first OIC method outperforming JPEG by using positional encodings \cite{mildenhall2021nerf} instead of raw coordinates as the input to a COIN-style MLP.

Ladune et al. identified that COIN's relatively low performance is due to the non-local nature of the MLP, i.e. each MLP parameter affects every pixel of the image and vice versa. To address this issue, they introduced Cool-Chic \cite{ladune2023cool}, which extends COIN with a hierarchical latent representation that contains most of the image information. The latent is entropy encoded using a neural autoregressive context model. After entropy decoding, the latent is upsampled such that each upsampled hierarchical layer of it has the same spatial resolution and is fed into a neural synthesis network which outputs the RGB values for each pixel. The entire codec is overfitted to a single image instance followed by quantization and entropy-encoding of the hierarchical latent, context and synthesis model. The Cool-Chic decoder is less complex compared to all LIC methods cited in the previous chapter, significantly outperforms \cite{dupont2021coin, dupont2022coinpp, strumpler2022implicit} and has a comparable RD-performance to the early LIC method \cite{balle2018variational}.

Cool-Chic has undergone iterative improvements: Leguay et al. \cite{leguay2023low} replaced Cool-Chic's fixed bicubic upsampling filter with a learnable one, replaced the synthesis MLP with a lightweight CNN and perform actual quantization of the latent during training using a straight-through estimator. C3 \cite{kim2024c3} added a latent resolution dependent context model, split the training into two stages with different quantization approximations and made small architectural changes. Blard et al. \cite{blard2024overfitted} added a CPU-only decoder as well as encoder speed-ups. Phillipe et al. \cite{philippe2024upsampling} made the adaptive upsampling filter separable and symmetric and added an additional filter branch leading to an improved performance. Ballé et al. \cite{balle2025good} use a Wasserstein
distortion-based loss to optimize C3 for human perception rather than pixel fidelity and extend the latent with a common randomness known without signaling to the en- and decoder to aid texture reproduction. In addition to these improvements, Cool-Chic was extended to support video \cite{leguay2024cool,leguay2025improved, ladune2025efficient, li2025cnvc}, lossless image coding \cite{zhang2025fitted} and scalable image coding \cite{benjak2025scalable, benjak2025progressive}. Zhang et al. \cite{zhang2026bridging} combine the concept of Cool-Chic with 2D Gaussian splatting by mapping overfitted latent maps to pixel values using a pixel-aligned learned rendering function.

MoRIC \cite{MoRIC2025} combines the pixel coordinate mapping of COIN \cite{dupont2021coin} with the learned latent of Cool-Chic \cite{ladune2023cool} and extends this joint architecture with multiple separate local synthesis networks (LSN) for distinct spatial regions that are each specifically tailored to their region's local distribution. As in COIN, the LSNs map pixel coordinates to RGB values, however, their intermediate features are modulated with the output of a shared global modulation network (GMN). As in  Cool-Chic, MoRIC uses an autoregressive context model to entropy encode a quantized hierarchical latent pyramid that is then upsampled and used as the input of the GMN. MoRIC is somewhat similar to our LANCE method in that both adapt to local changes in the image statistics. LANCE, however, has several distinct differences to MoRIC: LANCE adapts the context model for the entropy coder while MoRIC adapts by training separate synthesis models. LANCE implicitly learns a spatial hyperprior that is not limited in its expressiveness without any external input. Contrary to that, MoRIC uses static user-provided segmentation maps that are limited to 3 distinct regions.

\section{Proposed Method}
\label{method}
\begin{figure*}[tp]
	\centering
	\includegraphics[width=1\textwidth]{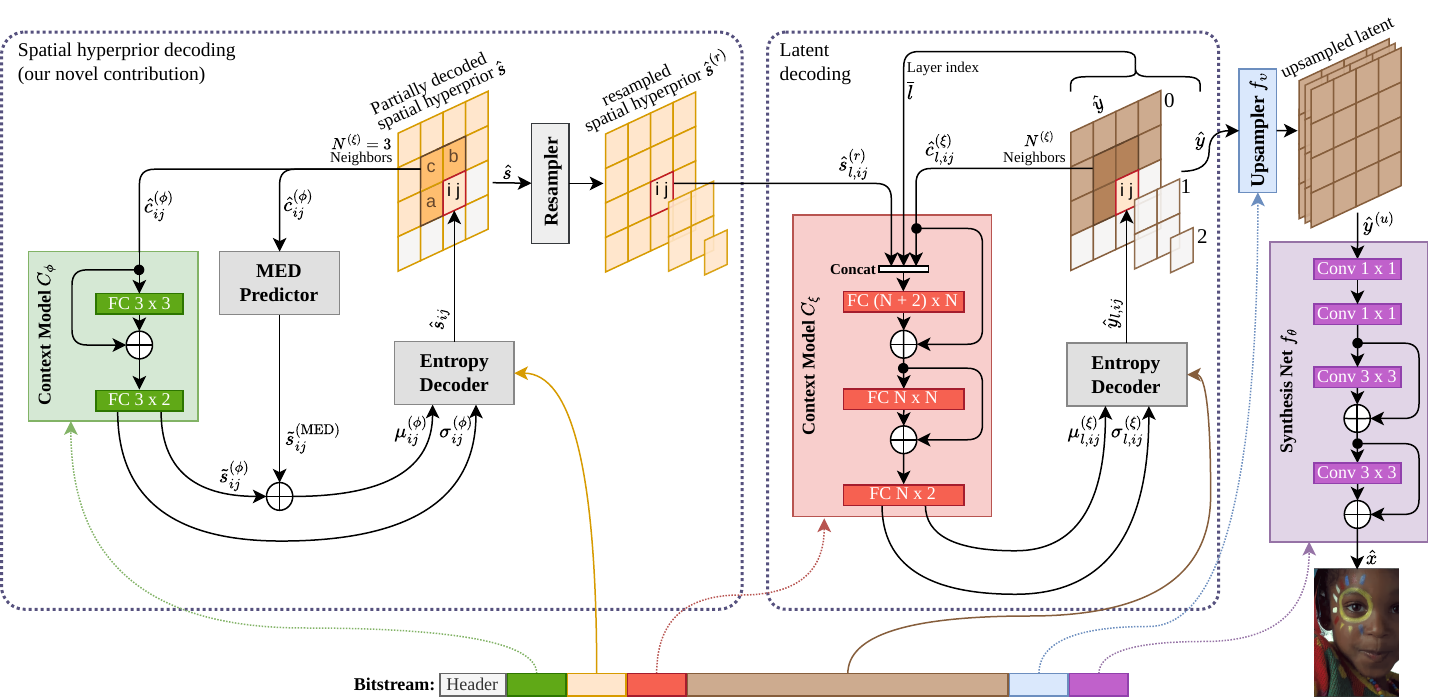}
	\caption{Decoding process of LANCE: First the spatial hyperprior $\bm{\hat{s}}$ is decoded in an autoregressive fashion using the context model $C_\phi$ and a Median Edge Detector (MED) predictor. The spatial hyperprior is then resampled into a pyramid of same shape and dimensionality as the latent $\bm{\hat{y}}$. The co-located element $\hat{s}^{(r)}_{l,ij}$ from the spatial hyperprior is then fed into the context model $C_\xi$ together with the normalized layer index $\bar{l}$ and the already decoded causal context $\bm{\hat{c}}^{(\xi)}_{l,ij}$ to decode the latent element $\hat{y}_{l,ij}$. Finally, the latent pyramid $\bm{\hat{y}}$ is upsampled using $f_\upsilon$ and fed into the synthesis model $f_\theta$ to generate the decoded image $\hat{x}$. All parameters are encoded in the bitstream. Parameter decoding is indicated by dotted lines from the bitstream to the parametrized modules. Gray blocks are static and do not contain any signaled parameters.}
	\label{fig:system_overview}
\end{figure*}
Existing overfitted image codecs rely on lightweight neural networks for entropy modeling which can only be adapted globally by signaling their parameters. Our codec represents a novel extension of the context model of Cool-Chic 4.0 \cite{leguay2025improved}, specifically designed to handle non-stationary image statistics. The core contribution of this work is Locally Adaptive Neural Context Estimation (LANCE), an additional stage of the entropy modeling process lifting the context-adaptive autoregressive model to adapt itself to local changes in image statistics. Because our method follows the instance-adaptive paradigm, where the decoder is tailored to a single image through gradient descent, we describe the system by guiding through the decoding steps in Section \ref{sec:decoding} followed by the gradient descent-based encoding procedure in Section \ref{sec:encoding}. Finally, a detailed description of the LANCE-integrated context model is provided in Section \ref{sec:lance_core}.

\subsection{Decoding Process and System Overview}
\label{sec:decoding}
A bitstream is processed following the architecture shown in Fig.~\ref{fig:system_overview}. The compressed information of an $H\times W$  pixel image is contained in three components: a hierarchical $L$-layer latent pyramid $\bm{\hat{y}} = \{\bm{\hat{y}}_{i} \in \mathbb{Z}^{H/2^i \times W/2^i},i=0, \cdots,L-1 \}$, a spatial hyperprior $\bm{\hat{s}} \in \mathbb{Z}^{H/2^d \times W/2^d}$ with downsampling exponent $d$, and the quantized parameters $\{\phi,\xi,\upsilon,\theta\}$ of the overfitted decoder modules. The decoding for an image proceeds as follows:

\textbf{Parameter Decoding:} The exponentially Golomb-encoded parameters $\phi$, $\xi$, $\upsilon$ and $\theta$ for the overfitted modules $C_\phi$, $C_\xi$, $f_\upsilon$ and $f_\theta$ are decoded.

\textbf{Adaptive Context Estimation:} This stage is divided into three sequential steps. First, the spatial hyperprior $\bm{\hat{s}}$, representing the core of LANCE, is entropy decoded in an autoregressive fashion using a static Median Edge Detector (MED) predictor \cite{weinberger1996loco, weinberger2000loco} and the context model $C_\phi$. Afterwards, $\bm{\hat{s}}$ is bicubically resampled into an $L$-layer hierarchical pyramid $\bm{\hat{s}}^{(r)}$ of the same dimensionality as the latent $\bm{\hat{y}}$.  Finally, the co-located element $\hat{s}^{(r)}_{l,ij}$, alongside the normalized layer index $\bar{l}$ as introduced in C3 \cite{kim2024c3}, is used to condition the subsequent context model $C_\xi$ to the local image statistics of the latent element $\hat{y}_{l,ij}$. The architecture of $C_\xi$ is similar to Cool-Chic 4.0 \cite{leguay2025improved} but is extended to accept the additional spatial hyperprior and layer index inputs to better handle non-stationary image statistics. $C_\xi$ predicts the distribution parameters $(\mu^{(\xi)}_{l,ij}, \sigma^{(\xi)}_{l,ij})$ used to entropy decode $\hat{y}_{l,ij}$.

LANCE does not introduce dependencies between the layers of $\bm{\hat{y}}$, allowing for parallel decoding of the layers.

\textbf{Latent Upsampling:} The layers of the decoded latent pyramid $\bm{\hat{y}}$ are upsampled to a tensor of shape $L \times H \times W$ using a cascade of learned $\times 2$ upsamplers $f_\upsilon$ as in \cite{philippe2024upsampling}.

\textbf{Synthesis Transform:} The upsampled latent tensor $\bm{\hat{y}}^{(u)}$ is processed by the synthesis network $f_\theta$ to reconstruct the image $\bm{\hat{x}}$. As in \cite{leguay2025improved}, the synthesis network consists of a sequence of 4 lightweight convolutional layers: two $1 \times 1$ convolution layers followed by two $3 \times 3$ convolution layers with residual connections.

\subsection{Encoding Process and Rate-Distortion Optimization}
\label{sec:encoding}
An image $x$ is encoded through gradient descent by jointly learning the latent $\bm{\hat{y}}$, spatial hyperprior $\bm{\hat{s}}$ and the parameters $\{\phi, \xi, \upsilon, \theta\}$ by minimizing the loss function
\begin{equation}
    \operatorname*{argmin}_{\{\phi, \xi, \upsilon, \theta\}, \bm{\hat{y}}, \bm{\hat{s}}} \mathrm{D}(x, \bm{\hat{x}}) + \lambda (\mathrm{R}(\bm{\hat{y}}) + \mathrm{R}(\bm{\hat{s}}))
    \label{loss}
\end{equation}
for a single instance of the image $x$, thus the parameters and latents are overfitted to this image. The Lagrange multiplier $\lambda\in\mathbb{R}$ balances the tradeoff between the rate $\mathrm{R}$ and the MSE distortion $\mathrm{D}$.

Quantization is the key element of any lossy image codec. We quantize the latent $\bm{\hat{y}}$ and spatial hyperprior $\bm{\hat{s}}$ using 
\begin{equation}
    Q(v) = \lfloor v\rceil,
    \label{quant}
\end{equation}
which, however, is not differentiable. Because of this, we approximate the quantization in a differentiable manner as in \cite{leguay2025improved} through
\begin{equation}
    Q_s(x, t) = \mathrm{softround}(\mathrm{softround}(x, t) + n, t)
\end{equation}
with the gaussian noise $n\sim \mathcal{N}$ and the soft rounding \cite{agustsson2020universally} function
\begin{equation}
    \mathrm{softround}(x, t) = \lfloor x \rfloor +
        \frac{\mathrm{tanh}(\frac{\Delta}{t})}{2\ \mathrm{\mathrm{tanh}(\frac{1}{2t})}}
        + \frac{1}{2},
\end{equation}
with $\Delta = x - \lfloor x \rfloor - 1/2$ and the soft rounding temperature $t$. During training, $t$ follows a decreasing schedule from $0.3$ down to $0.1$. To reduce the discrepancy of this approximation, the quantization is approximated using a straight-through estimator \cite{theis2017lossy} for the last iterations as introduced in \cite{leguay2023low}, which performs actual quantization in the forward pass while setting the partial gradient to $1$ for the backward pass.

The resulting parameters $\{\phi, \xi, \upsilon, \theta\}$, latent $\bm{\hat{y}}$ and spatial hyperprior $\bm{\hat{s}}$ form a compressed representation of the encoded image $\bm{\hat{x}}$. However, they have to be quantized and entropy encoded to enable efficient transmission, requiring an estimate $p$ of their unknown probability distribution $q$. The parameters $\{\phi, \xi, \upsilon, \theta\}$ contribute only little to the overall rate. Therefore, their rate is not modeled in the loss function in equation~\eqref{loss} and we entropy encode them after the training process and quantization using an exponential Golomb code assuming that their absolute values follow a geometric distribution.

The latent $\bm{\hat{y}}$ contributes most to the overall rate, because of which its rate $\mathrm{R}(\bm{\hat{y}})$ is optimized as part of the rate-distortion optimization in equation~\eqref{loss}. Even though the spatial hyperprior $\bm{\hat{s}}$ contributes little to the overall rate compared to $\bm{\hat{y}}$, including its rate $\mathrm{R}(\bm{\hat{s}})$ in equation~\eqref{loss} is necessary to limit the amount and type of information stored in $\bm{\hat{s}}$ to what is really needed. $\mathrm{R}(\bm{\hat{y}})$ and $\mathrm{R}(\bm{\hat{s}})$ are estimated by calculating the cross-entropy
\begin{equation}
    \mathrm{R}(v) = \mathrm{H}(v;q;p) = \mathbb{E}_{v \sim q} [-\log_2 p(v)],
    \label{crossentropy}
\end{equation}
which measures the average number of bits required to entropy encode a signal $v$ with the assumed distribution $p$ given the true distribution $q$. Real-world entropy coder implementations typically reach rates close to this.

\subsection{Context Modeling using LANCE}\label{sec:lance_core}
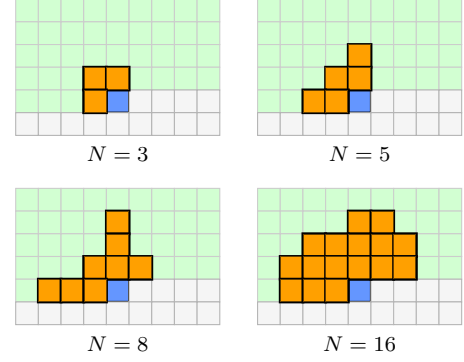
\begin{figure}[!t]
    \centering
    % No resizebox = no tiny text!
    \begin{tikzpicture}
        % Row 1
        \node (n3) at (0, 0) {\drawNTapFilter{0.3}{3}};
        \node (n5) at (3.2, 0) {\drawNTapFilter{0.3}{5}};
        
        \node[below=-3pt] at (n3.south) {\footnotesize $N=3$};
        \node[below=-3pt] at (n5.south) {\footnotesize $N=5$};
        
        % Row 2 (Tighter vertical jump at -2.8)
        \node (n8)  at (0, -2.5) {\drawNTapFilter{0.3}{8}};
        \node (n16) at (3.2, -2.5) {\drawNTapFilter{0.3}{16}};
        
        \node[below=-3pt] at (n8.south) {\footnotesize $N=8$};
        \node[below=-3pt] at (n16.south) {\footnotesize $N=16$};
    \end{tikzpicture}
    \caption{Visualization of the used $N$-tap context window configurations for the context models $C_\xi$ and $C_\phi$. The blue element is currently being en- or decoded. Gray elements are still unknown to the decoder-side.}
    \label{fig:filter_taps}
\end{figure}
To understand why context modeling is necessary, we can look at equation~\eqref{crossentropy} from another view:
\begin{equation}
\begin{aligned}
    \mathrm{H}(v;q;p)  = D_\text{KL}(q\parallel p) + \mathrm{H}(v).
\end{aligned}
\end{equation}
By this, we can see that $\mathrm{H}(v;q;p)$ can be minimized by reducing the entropy $\mathrm{H}(v)$ and by reducing the Kullback-Leibler divergence $D_\text{KL}(q\parallel p)$. The entropy $\mathrm{H}(v)$ can only be reduced by reducing the information content of $v$. We achieve this by quantizing the latent $\bm{\hat{y}}$ and spatial hyperprior $\bm{\hat{s}}$ using equation~\eqref{quant}. The amount of information that is lost, is controlled by the Lagrange multiplier $\lambda$ in equation~\eqref{loss} and causes the distortion $\mathrm{D}(\bm{x}, \bm{\hat{x}})$. Reducing the Kullback-Leibler divergence $D_\text{KL}(q\parallel p)$ means that $p$ has to be estimated as close to the true but unknown probability distribution $q$ of $v$ as possible. When $p$ equals $q$, $D_\text{KL}(q\parallel p)$ is reduced to zero resulting in the ideal case where we are able to entropy encode our signal at its entropy $\mathrm{H}(v)$.

Cool-Chic 4.0 \cite{leguay2025improved} estimates the probability distribution $p(\bm{\hat{y}})$ of the latent $\bm{\hat{y}}$ using the factorized model 
\begin{equation}
    p(\bm{\hat{y}}) = \prod_{l, ij} p(\hat{y}_{l,ij}\mid \psi; \bm{\hat{c}}_{l,ij}),
\end{equation}
where $l,ij$ addresses the element at position $(i,j)$ in the $l$-th layer of $\bm{\hat{y}}$, $\psi$ denotes the parameters of a learned autoregressive context model $C_\psi$ and $\bm{\hat{c}}_{l,ij}$ represents the $N$ already decoded neighbors of $\hat{y}_{l,ij}$. The probability of the latent element $\hat{y}_{l,ij}$ is calculated by integrating a parametrized Laplace distribution:
\begin{equation}
    p(\hat{y}_{l,ij}\mid \psi; \bm{\hat{c}}_{l,ij}) = \int_{\hat{y}_{l,ij}-0.5}^{\hat{y}_{l,ij}+0.5} g(y) \,dy,
    \label{eq:laplace_dist}
\end{equation}
with $g\sim \mathcal{L}(\mu_{l,ij}, \sigma_{l,ij})$ and $(\mu_{l,ij}, \sigma_{l,ij})=C_\psi(\bm{\hat{c}}_{l,ij})$. The context model $C_\psi$ therefore learns to map the context $\bm{\hat{c}}_{l,ij}$ to expectation and scale parameters of $g$. Please note that the last layer of $C_\psi$ outputs $\mu_{l,ij}$ and $\ln{\sigma_{l,ij}}$.

Conceptually, the autoregressive context model $C_\psi$ of Cool-Chic 4.0 \cite{leguay2025improved} is backward adaptive through the mapping from context $\bm{\hat{c}}_{l,ij}$ to the probability distribution $p(\bm{\hat{y}})$ and forward adaptive through the signaling of the overfitted parameters $\psi$. This form of forward adaptation can only adapt to global image statistics, since the same parameters are used for all positions within the latent $\bm{\hat{y}}$. However, natural images are rarely homogeneous meaning that the latent $\bm{\hat{y}}$ should be modeled as a non-stationary random field, requiring the parameters of the context model to adapt locally to spatial variations in image statistics. The existing $C_\psi$ architecture is unable to fulfill this requirement because it is a shallow neural network consisting of only three layers, designed specifically to keep signaling costs and computational complexity low. Shallow networks lack the necessary expressivity to adjust on their own to locally varying image statistics, thus failing to capture complex, non-stationary behaviors.

To overcome this limitation we propose LANCE, which extends Cool-Chic's context model by introducing forward-adaptive side information in form of the spatial hyperprior $\bm{\hat{s}}^{(r)}$ (see Fig.~\ref{fig:system_overview}) that explicitly models local variations in image statistics. Our context model $C_\xi$ is subsequently modified to accept this side information as a conditioning input, mapping both the causal context $\bm{\hat{c}}^{(\xi)}_{l,ij}$ and the co-located spatial hyperprior element $\hat{s}^{(r)}_{l,ij}$ to the distribution parameters. Furthermore, we feed a normalized latent layer index $\bar{l} = l/L$ into $C_\xi$ to enable the model to learn statistical characteristics specific to different layers within the latent pyramid \cite{kim2024c3}.

Consequently, the probability distribution for our modified context model is defined as:
\begin{equation}
    p(\bm{\hat{y}}) = \prod_{l, ij} p(\hat{y}_{l,ij}\mid \xi; \bm{\hat{c}}^{(\xi)}_{l,ij}; \hat{s}^{(r)}_{l,ij};\bar{l}),
\end{equation}
where $p(\hat{y}_{l,ij}\mid \xi; \bm{\hat{c}}^{(\xi)}_{l,ij}; \hat{s}^{(r)}_{l,ij};\bar{l})$ is calculated by integrating a Laplace distribution as in equation~\eqref{eq:laplace_dist} parametrized by our LANCE-enabled context model $C_\xi$:
\begin{equation}
(\mu^{(\xi)}_{l,ij}, \sigma^{(\xi)}_{l,ij}) = C_\xi(\bm{\hat{c}}^{(\xi)}_{l,ij}, \hat{s}^{(r)}_{l,ij}, \bar{l}).
\end{equation}
By conditioning on the spatial hyperprior $\hat{s}^{(r)}_{l,ij}$ and the normalized layer index $\bar{l}$, $C_\xi$ can dynamically adjust the distribution parameters to match the non-stationary statistics of the latent $\bm{\hat{y}}$ at different resolutions.

To implement $C_\xi$, we maintain the core architecture of the original Cool-Chic 4.0 context model $C_\psi$ while adapting it to accept our new inputs. Cool-Chic's context model $C_\psi$ consists of three fully connected layers, where the first two layers have a residual connection and map $N^{(\xi)}$ (one for each causal neighbor in $\bm{\hat{c}}^{(\xi)}_{l,ij}$) inputs to $N^{(\xi)}$ outputs, and the final layer maps to the two probability parameters ($\mu^{(\xi)}_{l,ij}, \sigma^{(\xi)}_{l,ij}$). The first layer is modified to accept a concatenated input vector consisting of the $N^{(\xi)}$ context neighbors $\bm{\hat{c}}^{(\xi)}_{l,ij}$, the co-located spatial hyperprior element $\hat{s}^{(r)}_{l,ij}$, and the normalized layer index $\bar{l}$, resulting in $N^{(\xi)}+2$ total inputs. To maintain high expressivity for the local context while minimizing complexity and signaling overhead, the residual connection within the network is applied only to the original $N^{(\xi)}$ context neighbor inputs, excluding the two new conditioning inputs. This limits the number of additional parameters in $\xi$ caused by our extension to $2N^{(\xi)}$ compared to $8N^{(\xi)}+16$ that would be necessary if the residual connection included the two new inputs.

Signaling the full-resolution spatial hyperprior $\bm{\hat{s}}^{(r)}$ directly is infeasible due to the prohibitive bitrate cost, as it has the same dimensionality as the latent $\bm{\hat{y}}$. To keep both the signaling and computational overhead minimal, we instead learn a single-layer spatial hyperprior $\bm{\hat{s}}$ with a significantly smaller spatial dimension than the final image. 

The compact hyperprior $\bm{\hat{s}}$ is resampled to match the dimensions of each layer $l$ of $\bm{\hat{y}}$ using a bicubic interpolation filter, which performs the necessary upsampling or downsampling operations depending on the target layer's resolution. 

In typical LIC schemes, the hyperprior is transform-coded using a dedicated hyper-encoder and decoder to de-correlate spatial dependencies, followed by entropy coding with a static, non-parametric probability model \cite{balle2018variational, minnen2018joint, cheng2020learned}. However, applying this approach to our overfitted codec poses a significant challenge: the hyper-decoder itself would need to be signaled, resulting in a substantial bitrate overhead that would negate the gains provided by the spatial hyperprior. Using the existing context model $C_\xi$ to model the probability distribution of the latent and the spatial hyperprior would reduce the overall performance as well, because the statistics of $\bm{\hat{s}}$ and $\bm{\hat{y}}$ differ vastly due to their different uses.

LOCO-I \cite{weinberger1996loco,weinberger2000loco}, the algorithm utilized in JPEG-LS, offers an alternative with low computational complexity and low signaling cost by employing a static Median Edge Detector (MED) predictor followed by gradient based context selection. The MED predictor
\begin{equation}
    \tilde{s}^{(\text{MED})}_{ij} = 
    \begin{cases}
      \min(a,b), & \text{if } c \geq \max(a,b)\\
      \max(a,b), & \text{if } c \leq \min(a,b)\\
      a+b-c, & \text{otherwise},
    \end{cases}   
    \label{eq:med}
\end{equation}
uses a primitive edge detector to test for vertical or horizontal edges in  the left, top, and top-left neighbor pixels $a, b, c \in \bm{\hat{c}}^{(\phi)}_{ij}$. If no edge is detected, the predicted value is $a+b-c$. Essentially, the predictor continues edges, if it detects any, and continues planes otherwise.

To efficiently model the probability distribution of the residual signal $\bm{\hat{s}} - \bm{\tilde{s}}^{(\text{MED})}$, LOCO-I calculates local gradients from the neighbors ${a,b,c}$ that are individually quantized into a small set of values. The combination of these quantized gradients forms a context index, resulting in a large number of contexts (1094 in the original implementation). Context-specific statistics are tracked by counting the number of occurrences and accumulating residual magnitudes for each context. This scheme is fundamentally sequential, rendering it unfeasible during training of our overfitted codec. One could theoretically adapt this approach by learning (instead of accumulating) specific parameters for the different contexts in LOCO-I and signaling them forward, however, this would lead to an unreasonable bitrate overhead. Furthermore, the context selection mechanism in LOCO-I is based on hard decisions (thresholding gradients), which introduces non-differentiable operations that make end-to-end training unstable.

Our solution is a hybrid scheme where we combine the static and low-complexity MED predictor of LOCO-I with a small and lightweight learned context model $C_\phi$. As the MED predictor, $C_\phi$ has a context window $\bm{\hat{c}}^{(\phi)}_{ij}$ containing $N^{(\phi)}=3$ causal neighbors (see Fig.~\ref{fig:filter_taps}). The architecture of the context model is similar to $C_\psi$ used in Cool-Chic 4.0 \cite{leguay2025improved}, but with only one $3 \times 3$ fully connected layer with a residual connection followed by a ReLU activation function and another fully connected $3 \times 2$ layer (see Fig.~\ref{fig:system_overview}). The reduced number of layers combined with the low number of $N^{(\phi)}=3$ context elements results in $\phi$ containing only $20$ parameters. The probability distribution of $\bm{\hat{s}}$ is consequently modeled by
\begin{equation}
    p(\bm{\hat{s}}) = \prod_{ij} p(\hat{s}_{ij}\mid \phi; \bm{\hat{c}}^{(\phi)}_{ij}),
\end{equation}
where $p(\hat{s}_{ij}\mid \phi; \bm{\hat{c}}^{(\phi)}_{ij})$ is calculated by integrating a Laplace distribution as in equation \eqref{eq:laplace_dist} parametrized by
\begin{equation}
\mu^{(\phi)}_{ij} = \tilde{s}^{(\phi)}_{ij} + \tilde{s}^{(\text{MED})}_{ij}
\end{equation}
and
\begin{equation}
(\tilde{s}^{(\phi)}_{ij}, \sigma^{(\phi)}_{ij}) = C_\phi(\bm{\hat{c}}^{(\phi)}_{ij}).
\end{equation}

The MED predictor in equation \eqref{eq:med} consists of 3 distinct predictors that are selected using hard thresholds, rendering it non differentiable. Therefore, we introduce the differentiable approximation
\begin{equation}
    \begin{aligned}
        \tilde{s}^{(\text{MED})}_{ij} = &\min(a,b)\cdot s_{\text{max}} \\
        &+\max(a,b)\cdot s_{\text{min}} \\
        &+(a+b-c)(1-s_{\text{max}}-s_{\text{min}}),
    \end{aligned}
    \label{eq:meddiff}
\end{equation}
with $s_{\text{max}} = \sigma(\frac{c-\max(a,b)}{t})$ and $s_{\text{min}} = \sigma(\frac{\min(a,b)-c}{t})$, which replaces it during training. $\sigma(\cdot)$ represents the sigmoid function and the temperature $t$ controls the softness of the approximation. During training, $t$ follows a decreasing schedule from $0.3$ down to $0.1$ and for the last iterations of the training, $\bm{\tilde{s}}^{(\text{MED})}$ is approximated using a straight-through estimator \cite{theis2017lossy}.

Our method of encoding the spatial hyperprior offers substantial benefits. Because the MED predictor is static, it requires no signaling overhead. Furthermore, the limited number of layers and small context window of $C_\phi$ ensure that the computational complexity and signaling overhead remain low.

\section{Evaluation}
\label{evaluation}
This chapter evaluates LANCE by comparing its rate-distortion efficiency and decoder complexity against established benchmarks.
\subsection{Experiment Configuration}
\begin{figure}[!t]
    \centering
    \input{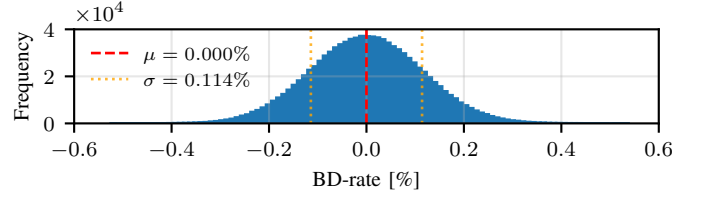}
    \caption{Histogram of the BD-rate between 1 million random permutations of per quality and image data points sampled from ten fully independent Cool-Chic 4.0 \cite{leguay2025improved} encodings with 5 quality levels versus the average PSNR and bitrate of the same ten encodings evaluated on the Kodak dataset \cite{kodak}.}
    \label{fig:hist}
  \end{figure}
    \begin{figure}[!t]
    \centering
    \input{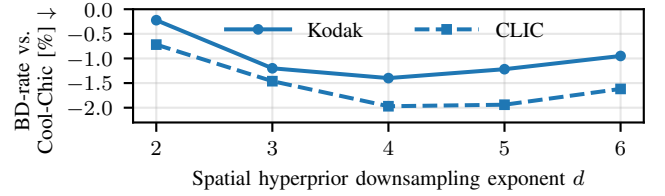}
    \caption{BD-rate of our method LANCE versus Cool-Chic 4.0 \cite{leguay2025improved} over the spatial prior downsampling factor $d$ evaluated on the Kodak \cite{kodak} and CLIC 2020 professional validation \cite{clic2020} datasets.}
    \label{fig:downfactor}
  \end{figure}
Our proposed codec is identical to Cool-Chic 4.0 \cite{leguay2025improved} besides the context modeling aspect which we extended with our LANCE method. In the following, we refer to our codec as LANCE.

We compare LANCE with Cool-Chic 4.0, which is a video extension of Cool-Chic 3.4 \cite{philippe2024upsampling} with both sharing the same intra codec architecture and the VVC \cite{bross2021developments_VVC} reference software VTM version 23.14. Additionally, we also compare to the overfitted codec MoRIC \cite{MoRIC2025} and to the autoencoder-based LIC methods MLIC++ \cite{jiang2025mlic} and Cheng 2020 \cite{cheng2020learned}.

Both Cool-Chic 4.0 and LANCE are, besides the changes that we introduced, configured in the same way to ensure a meaningful comparison. The latent $\bm{\hat{y}}$ consists of $L=7$ layers, the context model has two residual layers and the upsampler uses the filter configuration from \cite{philippe2024upsampling}. The quality operation points $\lambda\in\{0.0001, 0.0004, 0.001, 0.004, 0.02\}$ are the same for both as well. We evaluate both methods at the same three decoder complexity operation points following the parameters in Table~\ref{tab:operation_point_config} and trained both for 100,000 iterations. The spatial hyperprior of LANCE is configured with a downsampling exponent of $d=4$ and its context model $C_\phi$ has two layers and a context window size $N^{(\phi)}$ of three for all three operation points. We report rates for Cool-Chic and LANCE based on the full file size of the written bitstreams.

VTM is configured to use the "main 10 444" profile. This profile has a bit depth of ten bit and uses 4:4:4 chroma format. Since LANCE operates in RGB domain, we convert RGB images to YUV BT.709 before encoding with VTM and back to RGB after decoding.

MoRIC is evaluated at three different complexity operation points with fixed decoder architectures selected from the default configuration for \textit{standard image compression} (see Table 2 of the appendix in \cite{MoRIC2025}). These three operation points utilize 18, 18, and 24 synthesis features paired with ARM input window sizes of 8, 16, and 16, respectively. The MoRIC encoder software does not use an entropy coder to write actual bitstreams. Instead, it reports estimated rates. We modified the MoRIC software to use the same entropy coders for network parameters and latents as LANCE and Cool-Chic use and added the size of Cool-Chic's header to the total rate to ensure a fair comparison. We report results for the unmodified software which we refer to as \textit{MoRIC (estimated rate)} and for our modification \textit{MoRIC (real rate)}.

\begin{table}[h!]
    \centering
    \caption{Parameter configuration for the three operation points used for LANCE and Cool-Chic 4.0. }
    \label{tab:operation_point_config}
    \setlength{\tabcolsep}{3pt}
    \begin{tabular}{c|ccc}
    \toprule
    Parameter & HOP & MOP & LOP \\
    \midrule
    Context size $N^{(\xi)}$ & 16 & 16 & 8\\
    Synthesis channels& 48 & 16 & 16 \\
    \bottomrule
    \end{tabular}
\end{table} 

We evaluate on the Kodak \cite{kodak} dataset, which contains 24 $768\times512$ pixel images, and CLIC 2020 professional validation \cite{clic2020} dataset, which contains 41 images with resolutions ranging from $384\times512$ to $1370\times2048$ using PSNR as the quality metric (computed in RGB444 domain). BD-rates \cite{bjontegaard2001calculation} are calculated using Akima interpolation. 
\subsection{Variance of Cool-Chic}

Overfitted codecs such as Cool-Chic use stochastic gradient descent to encode images. The stochastic nature of the encoding process causes the encoding performance to vary when the same image with the same configuration is encoded multiple times. To get an understanding of the extend of this variance, we encoded the Kodak dataset ten times using Cool-Chic 4.0. This gives us $5\cdot24=120$ (5 quality operation points and 24 images) PSNR and bitrate value pairs for each of these ten encodings. Consequently, there are $10^{120}$ different possible permutations of PSNR and bitrate value pairs that we can sample from. We sampled one million random permutations and calculated the BD-rate for all of them versus the average per picture and quality operation point PSNR and bitrate of the same encodings which results in the histogram in Fig.~\ref{fig:hist}.

The close to zero mean of this data shows that the per image and quality operation point averaging of PSNR and bitrate values is valid. The variance of $0.114\%$ might appear low, however, the best possible BD-rate that we can obtain by selecting the 120 data points with the lowest evaluation loss from the ten encodings versus our average value is $-1.52 \%$. 

To ensure a fair and meaningful comparison, we repeated all experiments that we report in this paper five times for Cool-Chic 4.0 and LANCE and calculate BD-rates on per image and per quality operation point averaged PSNR and bitrate values over all five experiment runs.
\subsection{Optimal Parameters} 
Fig.~\ref{fig:downfactor} shows that the optimal value for the spatial hyperprior downsampling exponent is $d=4$ across both tested datasets. However, the curve is shifted towards higher $d$ for CLIC 2020, implying that higher resolution images can be encoded with higher $d$ values. Increasing the resolution leads to higher signaling cost for $\bm{\hat{s}}$ while decreasing the resolution limits its expressiveness.

The context model $C_\phi$ that is used jointly with the MED-predictor to compress $\bm{\hat{s}}$ has only one residual layer compared to two in $C_\xi$. Furthermore, $C_\phi$ has the same context window $\bm{\hat{c}}^{(\phi)}_{ij}$ of size $N^{(\phi)}=3$ as the MED-predictor, which is also smaller than that of $C_\xi$. Table~\ref{tab:context_parameters} shows that this configuration results in the best rate distortion performance. Larger model sizes lead to a higher parameter count that increases the signaling cost. 
  
\begin{table}[h!]
    \centering
    \caption{Comparison of parameter count $n_P$ and BD-rate (LANCE vs. Cool-Chic 4.0) for different number $n_L$ of residual layers and context window size $N^{(\phi)}$ of the spatial hyperprior context model $C_\phi$.}
    \label{tab:context_parameters}
    \begin{tabular}{c|cc|cc}
        \toprule
                & \multicolumn{2}{c|}{$N^{(\phi)}=3$}& \multicolumn{2}{c}{$N^{(\phi)}=5$} \\
                & BD-rate $\downarrow$ & $n_P$ & BD-rate $\downarrow$ & $n_P$ \\
        \midrule
        $n_L=1$ & \textbf{-1.40\%} & 20 & -1.34\% & 42\\
        $n_L=2$ & -1.32\% & 32 & -1.13\% & 72 \\
        \bottomrule
    \end{tabular}
\end{table}

\subsection{Rate-Distortion Performance and Complexity}
\label{sec:rate_complexety}
\begin{figure}[!t]
    \centering
    \input{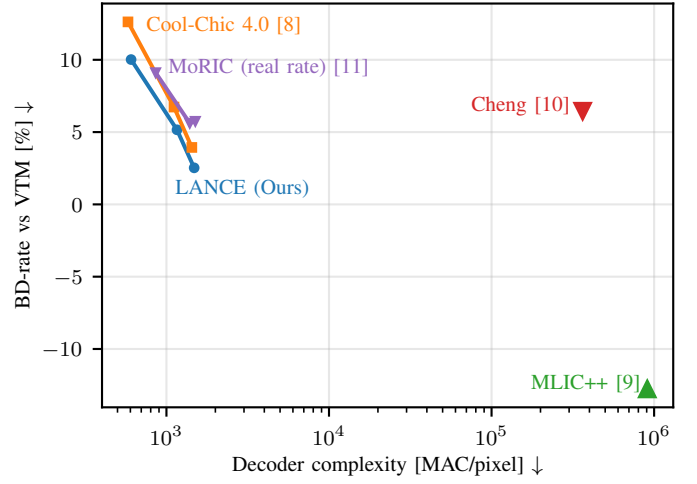}
    \caption{Rate savings in terms of BD-rate versus VVC (VTM 23.14) over decoder complexity evaluated on the Kodak dataset \cite{kodak}. Negative results: less rate is required to get the same quality as VVC.}
    \label{fig:complexity_kodak}
  \end{figure}

Fig.~\ref{fig:complexity_clic} and Fig.~\ref{fig:complexity_kodak} show the BD-rate over decoder complexity for all methods evaluated on the CLIC and Kodak datasets, respectively. Table~\ref{tab:bd_rates} and Table~\ref{tab:bd_rates_coolchic} show the corresponding BD-rate values versus VTM and Cool-Chic 4.0, respectively. LANCE has a better BD-rate to complexity tradeoff compared to Cool-Chic 4.0, i.e. LANCE has a lower decoder complexity at the same quality level and a better BD-rate at the same decoder complexity.

\begin{table}[h!]
    \centering
    \caption{BD-rates of all methods versus VVC (VTM 23.14) evaluated on the Kodak \cite{kodak} and CLIC 2020 professional validation \cite{clic2020} datasets.}
    \label{tab:bd_rates}
    \setlength{\tabcolsep}{2pt}
    \begin{tabular}{c|ccc|ccc}
    \toprule
    &\multicolumn{3}{c|}{Kodak}&\multicolumn{3}{c}{CLIC} \\
    Method& HOP & MOP & LOP & HOP & MOP & LOP \\
    \midrule
    LANCE (ours)& 2.53 & 5.16 & 10.02 & -2.42 & 0.21 & 4.63 \\
    Cool-Chic \cite{leguay2025improved}& 3.94 & 6.73 & 12.62 & -0.47 & 2.60 & 7.77\\
    MoRIC (real rate) \cite{MoRIC2025}& 5.75 & 5.62 & 9.11 & 13.47 & 4.24 & 16.52\\
    MoRIC (estimated rate) \cite{MoRIC2025}& 4.23 & 4.07 & 7.78& 11.81 & 2.77 & 15.27\\
    Cheng \cite{cheng2020learned}& \multicolumn{3}{c|}{6.44}  &\multicolumn{3}{c}{6.03} \\
    MLIC++ \cite{jiang2025mlic}&   \multicolumn{3}{c|}{-12.76}  &\multicolumn{3}{c}{-12.38}  \\

    \bottomrule
    \end{tabular}
\end{table} 
\begin{table}[h!]
    \centering
    \caption{ BD-rates of LANCE versus Cool-Chic 4.0 at similar decoder complexity operation points evaluated on the Kodak \cite{kodak} and CLIC 2020 professional validation \cite{clic2020} datasets.}
    \label{tab:bd_rates_coolchic}
    \setlength{\tabcolsep}{2pt}
    \begin{tabular}{c|ccc|ccc}
    \toprule
    &\multicolumn{3}{c|}{Kodak}&\multicolumn{3}{c}{CLIC} \\
    Method& HOP & MOP & LOP & HOP & MOP & LOP \\
    \midrule
    LANCE (ours)& -1.40 & -1.54 & -2.41 & -1.97 & -2.36 & -2.99 \\

    \bottomrule
    \end{tabular}
\end{table} 

LANCE also outperforms MoRIC, which uses multiple region dependent synthesis networks to adapt to local changes in image statistics. Please note that MoRIC loses 1.3\% to 1.5\% in terms of BD-rate when the rate is measured based on bitstream file sizes compared to estimated rates. Due to observed instabilities in MoRIC’s performance, we have limited its representation in Fig.~\ref{fig:complexity_clic} to ensure a fair comparison. All results across all tested complexity operation points are available in Table~\ref{tab:bd_rates}.
  
The older LIC method by Cheng et al. has a worse rate-distortion performance than LANCE and Cool-Chic 4.0 at orders of magnitude higher decoder complexity. MLIC++ has a even higher decoder complexity, but outperforms both LANCE and Cool-Chic 4.0 by more than 10\% in terms of BD-rate. This shows that LIC achieves superior compression performance, but OIC's significantly lower decoder complexity makes it far more practical for resource-constrained hardware. 

\begin{figure*}[!t]
    \centering
    \includegraphics[width=1\textwidth]{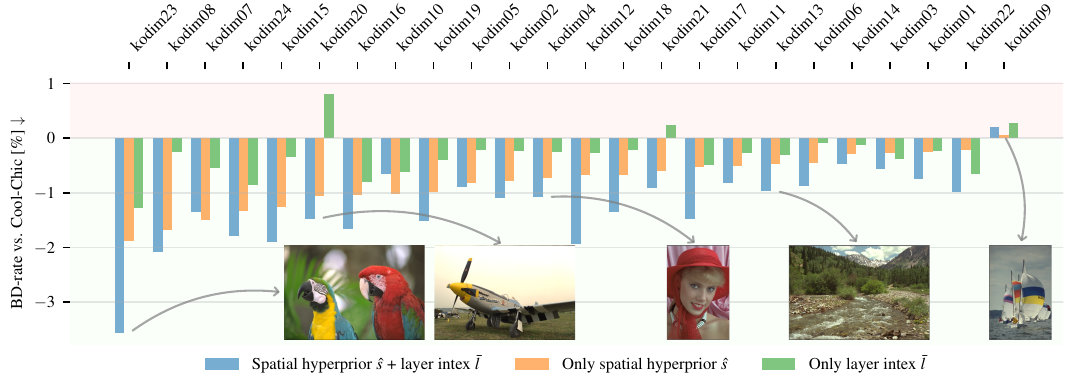}
    \caption{Per-image BD-rates of our method LANCE versus Cool-Chic 4.0 \cite{leguay2025improved} evaluated on the Kodak dataset \cite{kodak}. LANCE is tested with all its modules activated (blue), with deactivated spatial hyperprior (green) and with deactivated layer index (orange). The bars are sorted by the value of the orange bars.}
    \label{fig:per_image_performance}
  \end{figure*}

Fig.~\ref{fig:per_image_performance} shows the rate-distortion performance of LANCE separately for every image of the Kodak dataset. To better understand whether the gains come from the spatial hyperprior $\bm{\hat{s}}$ or the layer index $\bar{l}$, we evaluated three LANCE configurations, one with all modules activated, one with the layer index $\bar{l}$ deactivated, and one with the spatial hyperprior $\bm{\hat{s}}$ deactivated. LANCE outperforms Cool-Chic 4.0 in the first two cases for every image except one. The spatial hyperprior $\bm{\hat{s}}$ yields higher gains than the layer index $\bar{l}$ in most cases. The combination of $\bm{\hat{s}}$ and $\bar{l}$ performs better in most cases than $\bm{\hat{s}}$ and $\bar{l}$ by themselves.

\begin{figure*}[htbp]
\centering

\renewcommand{\arraystretch}{1.5} 

% Kodak Images
\begin{adjustbox}{width=\textwidth,center}
\scriptsize
\begin{tabular}{@{}l@{\hspace{-2pt}}c@{\hspace{2mm}}c@{\hspace{2mm}}c@{\hspace{2mm}}c@{\hspace{2mm}}c@{\hspace{2mm}}c@{}}
 & Original & \textbf{$\lambda=0.0001$} & \textbf{$\lambda=0.0004$} & \textbf{$\lambda=0.001$} & \textbf{$\lambda=0.004$} & \textbf{$\lambda=0.02$} \\

% kodim1
\raisebox{0.8cm}{\parbox[c]{2cm}{\centering Kodak\\kodim01\\BD-rate -0.74\% \\ $768\times512$}} & 
\includegraphics[width=2.5cm]{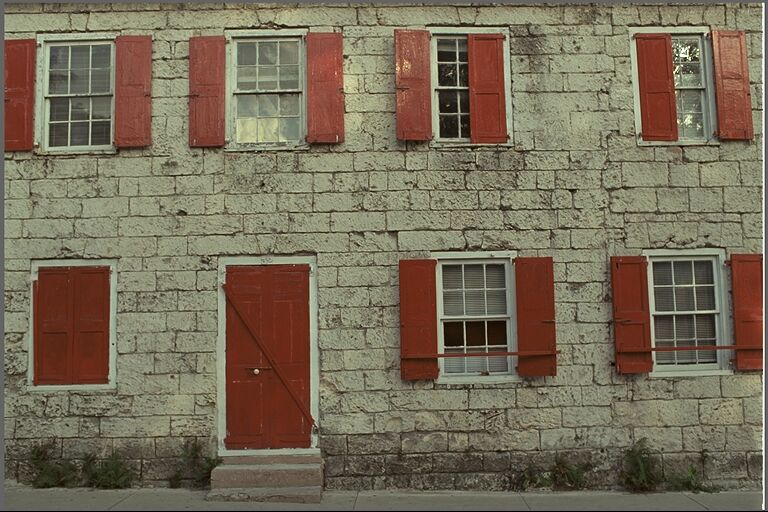} &
\includegraphics[width=2.5cm]{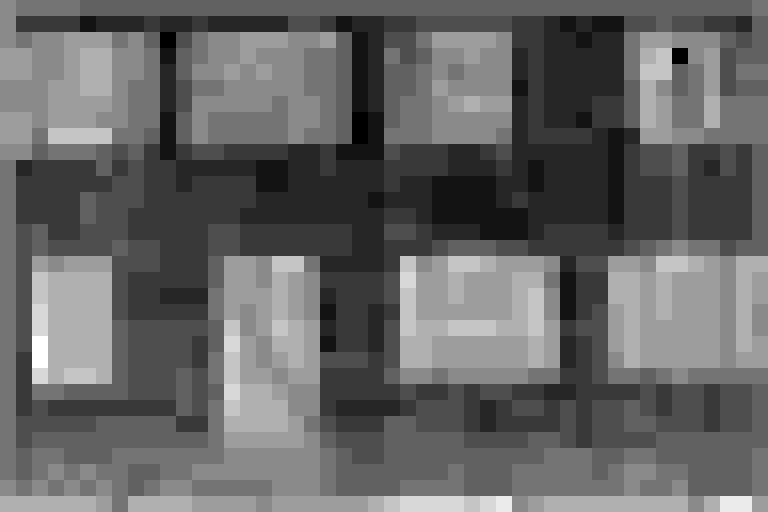}  &
\includegraphics[width=2.5cm]{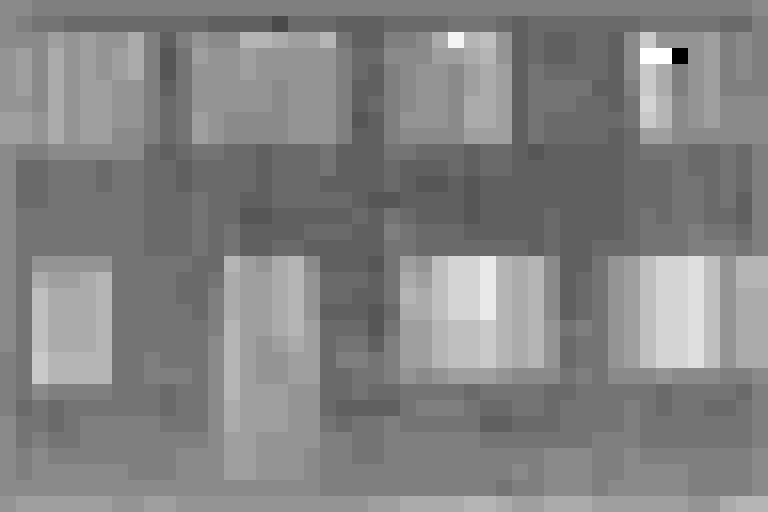}  &
\includegraphics[width=2.5cm]{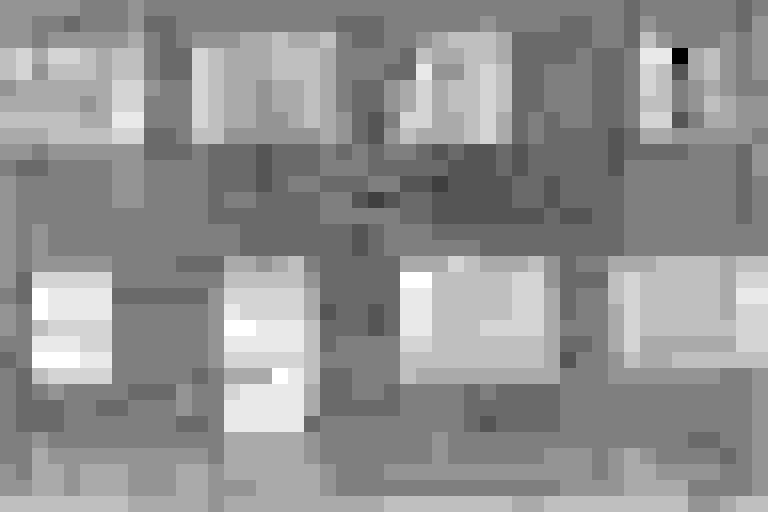}  &
\includegraphics[width=2.5cm]{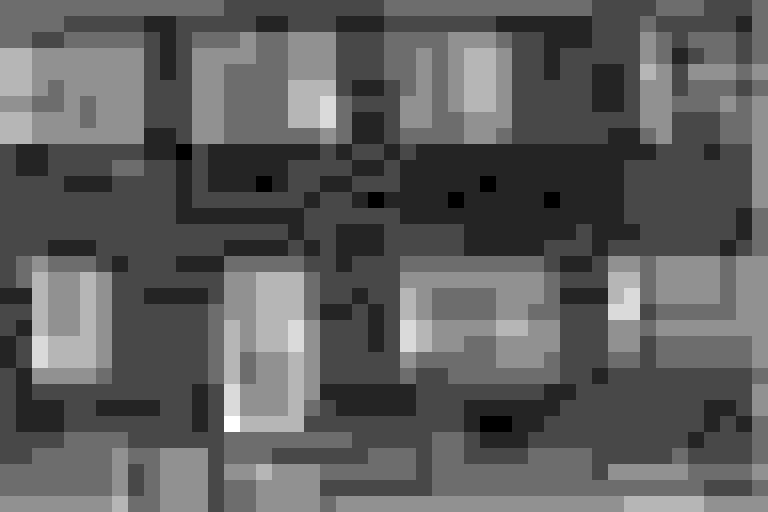}  &
\includegraphics[width=2.5cm]{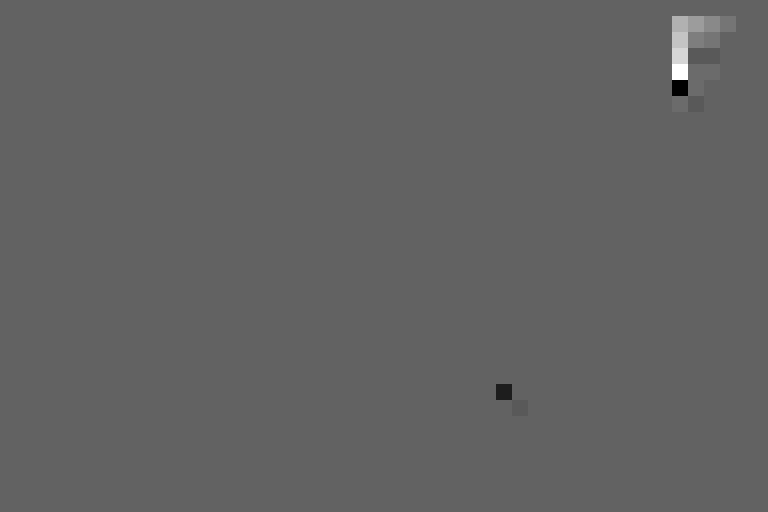}  \\

% kodim7
\raisebox{0.8cm}{\parbox[c]{2cm}{\centering Kodak\\kodim07\\BD-rate -1.35\% \\ $768\times512$}} & 
\includegraphics[width=2.5cm]{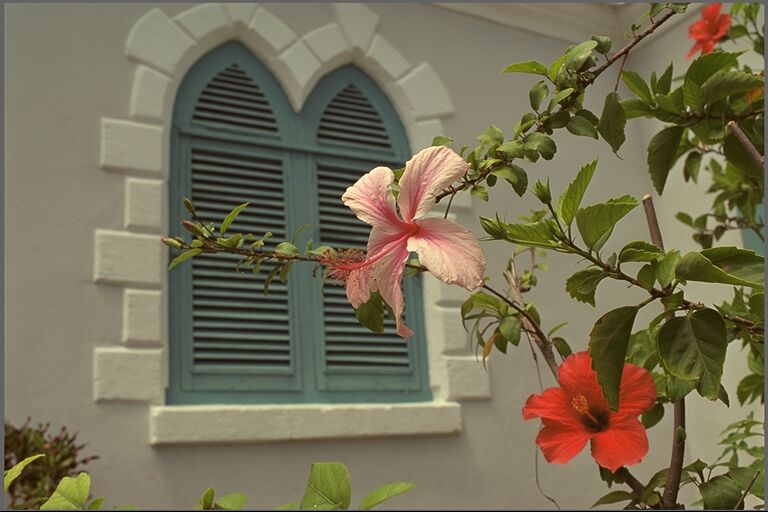} &
\includegraphics[width=2.5cm]{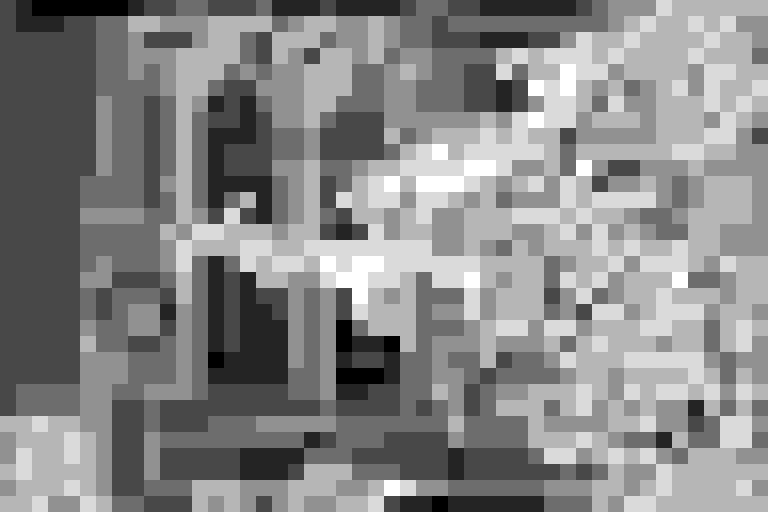}  &
\includegraphics[width=2.5cm]{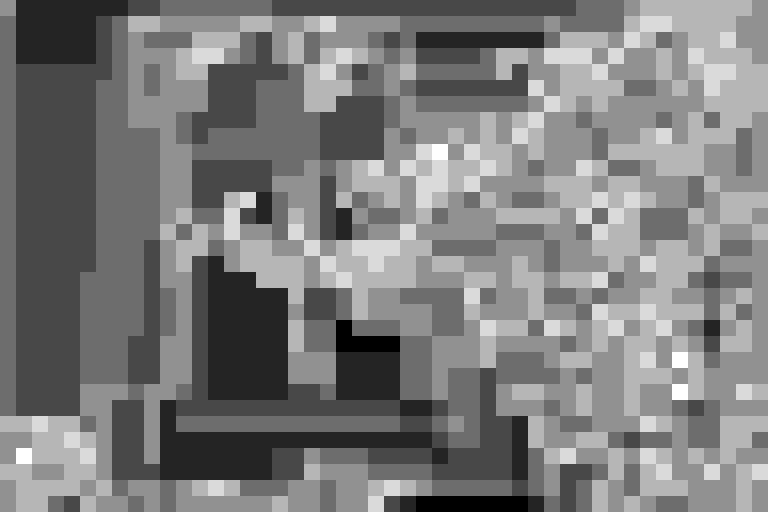}  &
\includegraphics[width=2.5cm]{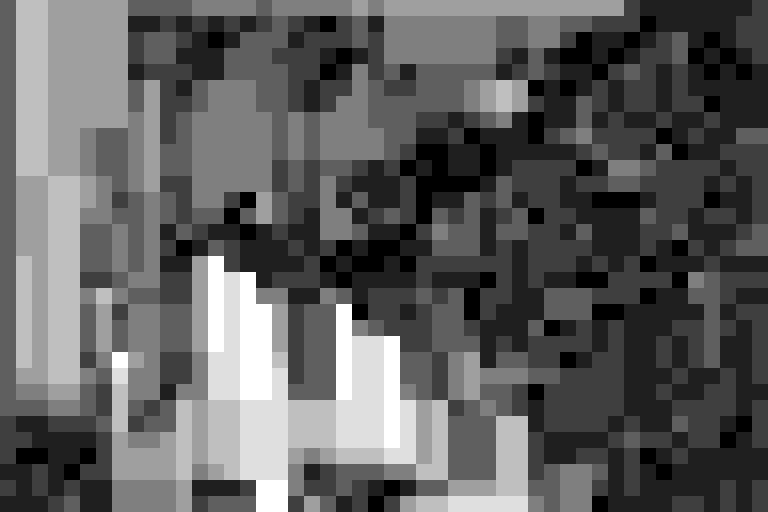}  &
\includegraphics[width=2.5cm]{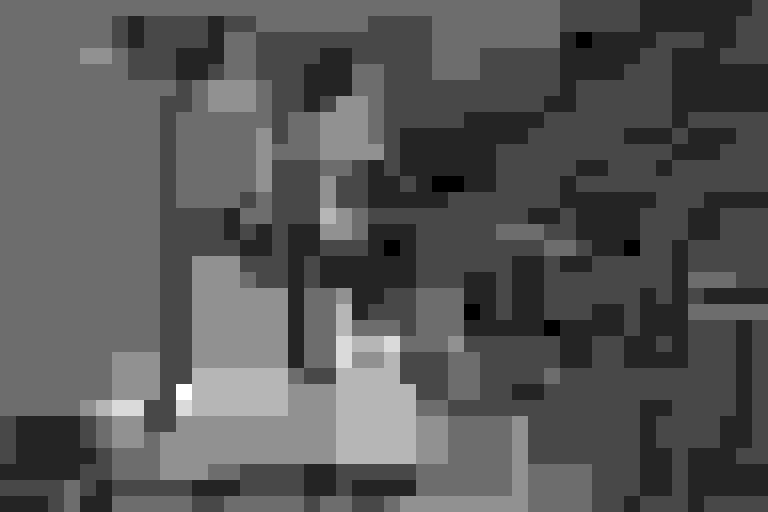}  &
\includegraphics[width=2.5cm]{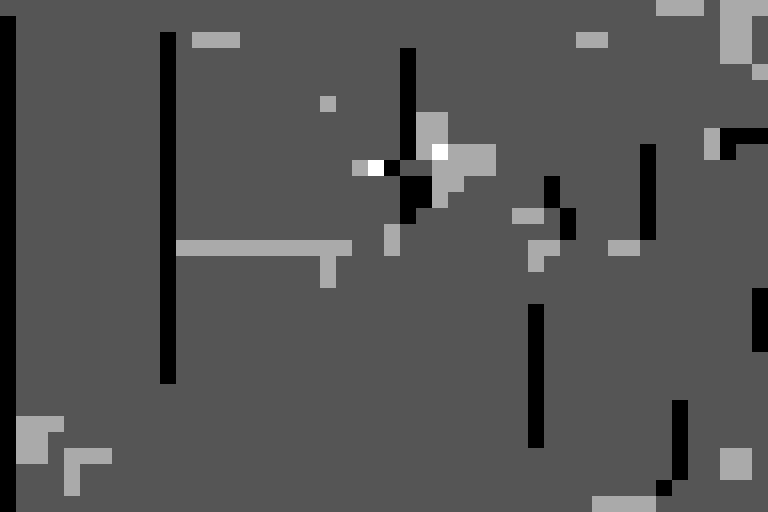}  \\

% kodim15
\raisebox{0.8cm}{\parbox[c]{2cm}{\centering Kodak\\kodim15\\BD-rate -1.90\% \\ $768\times512$}} & 
\includegraphics[width=2.5cm]{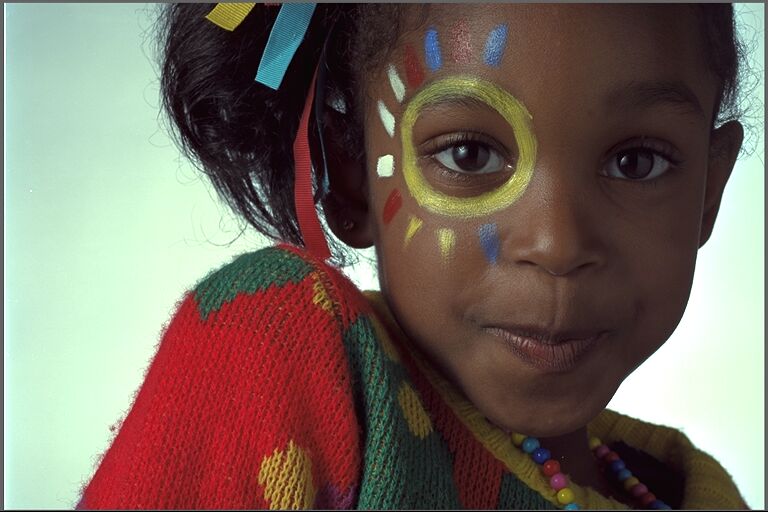} &
\includegraphics[width=2.5cm]{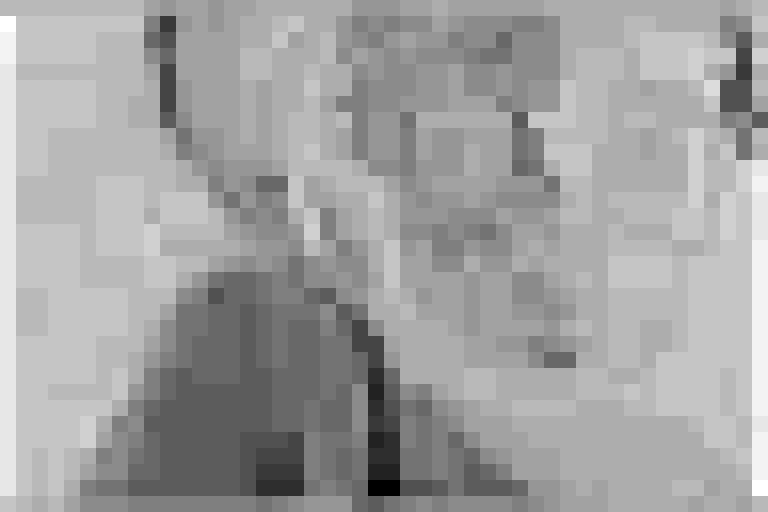}  &
\includegraphics[width=2.5cm]{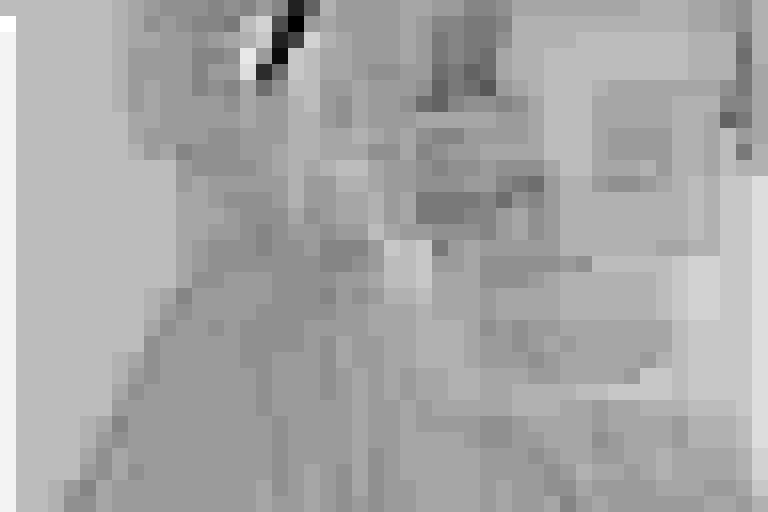}  &
\includegraphics[width=2.5cm]{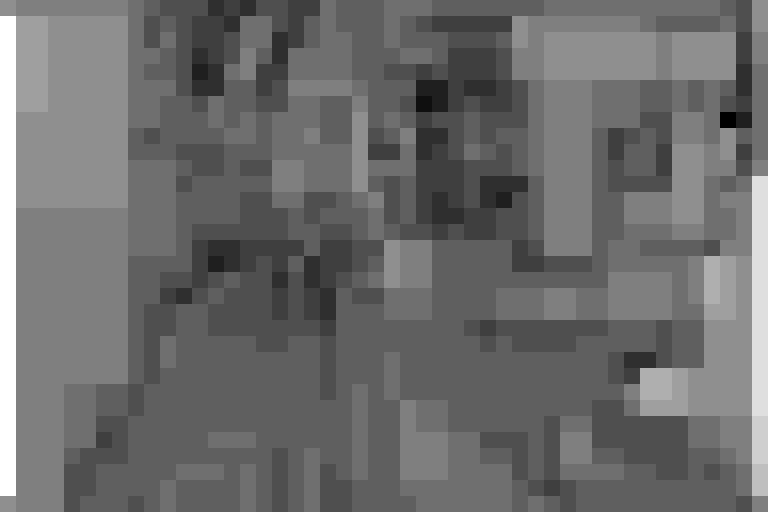}  &
\includegraphics[width=2.5cm]{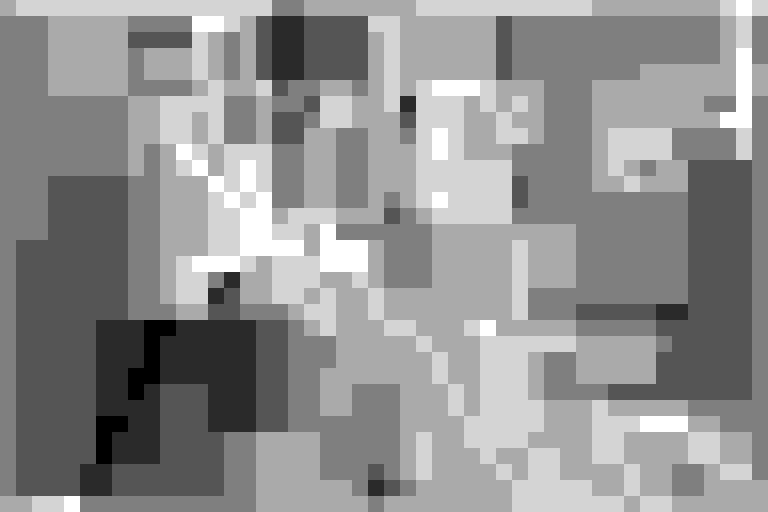}  &
\includegraphics[width=2.5cm]{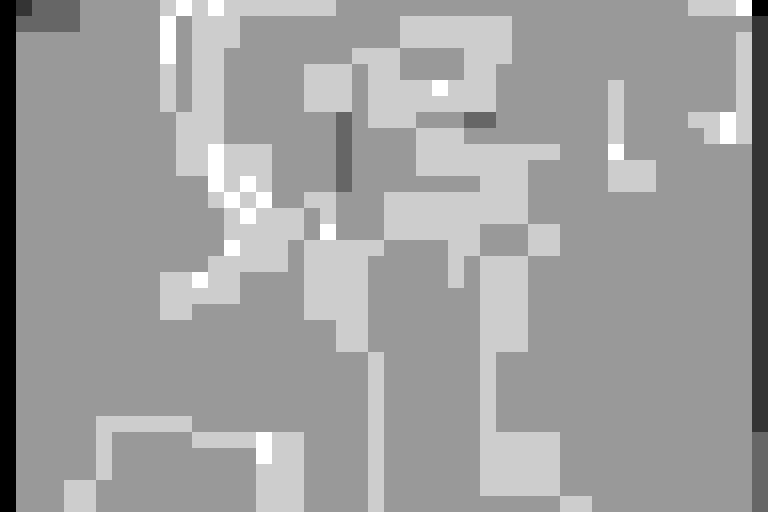}  \\

% jeremy-cai-1174
\raisebox{0.8cm}{\parbox[c]{2cm}{\centering CLIC\\jeremy-cai\\BD-rate -2.21\% \\ $1936\times1200$}} & 
\includegraphics[width=2.5cm]{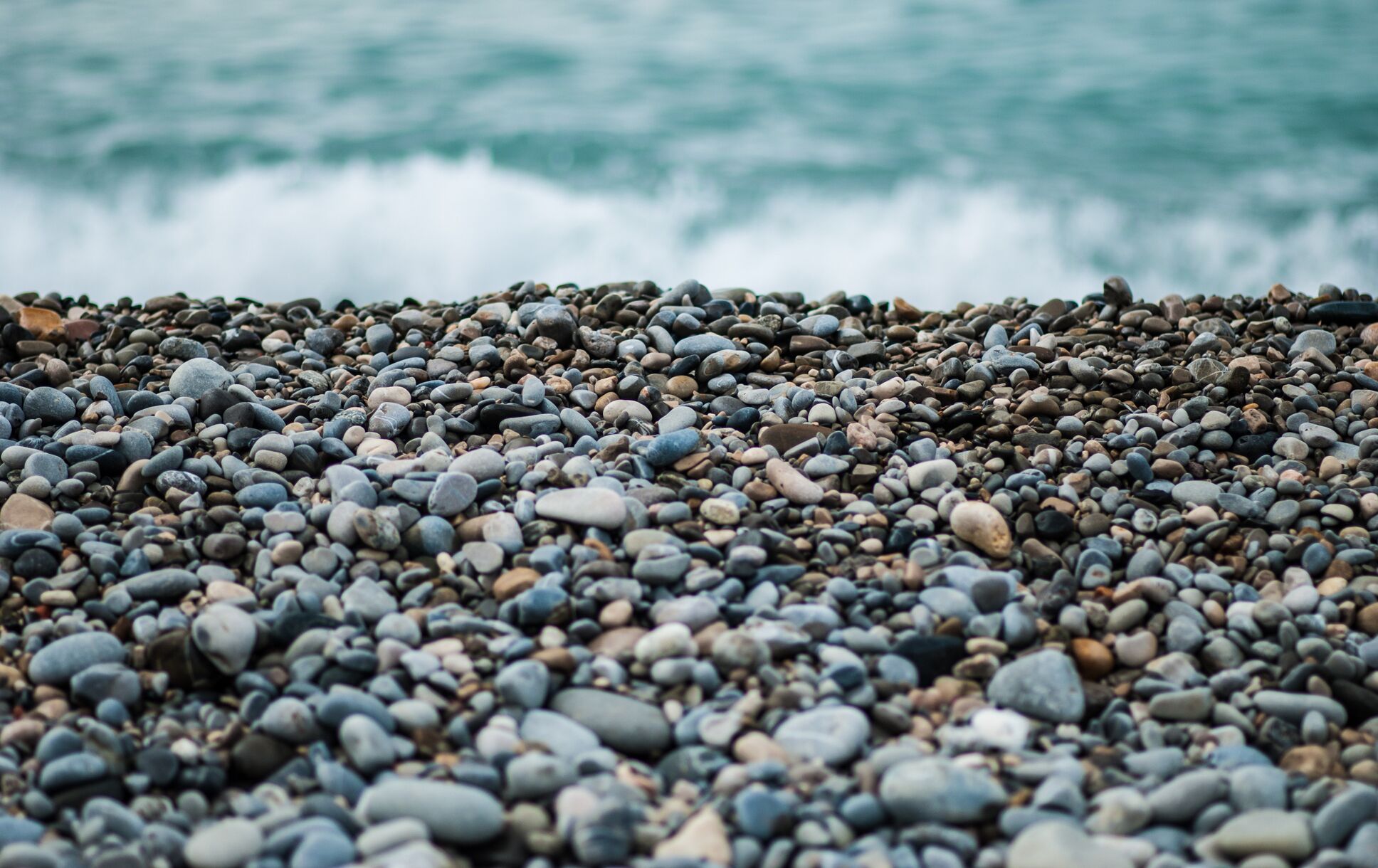} &
\includegraphics[width=2.5cm]{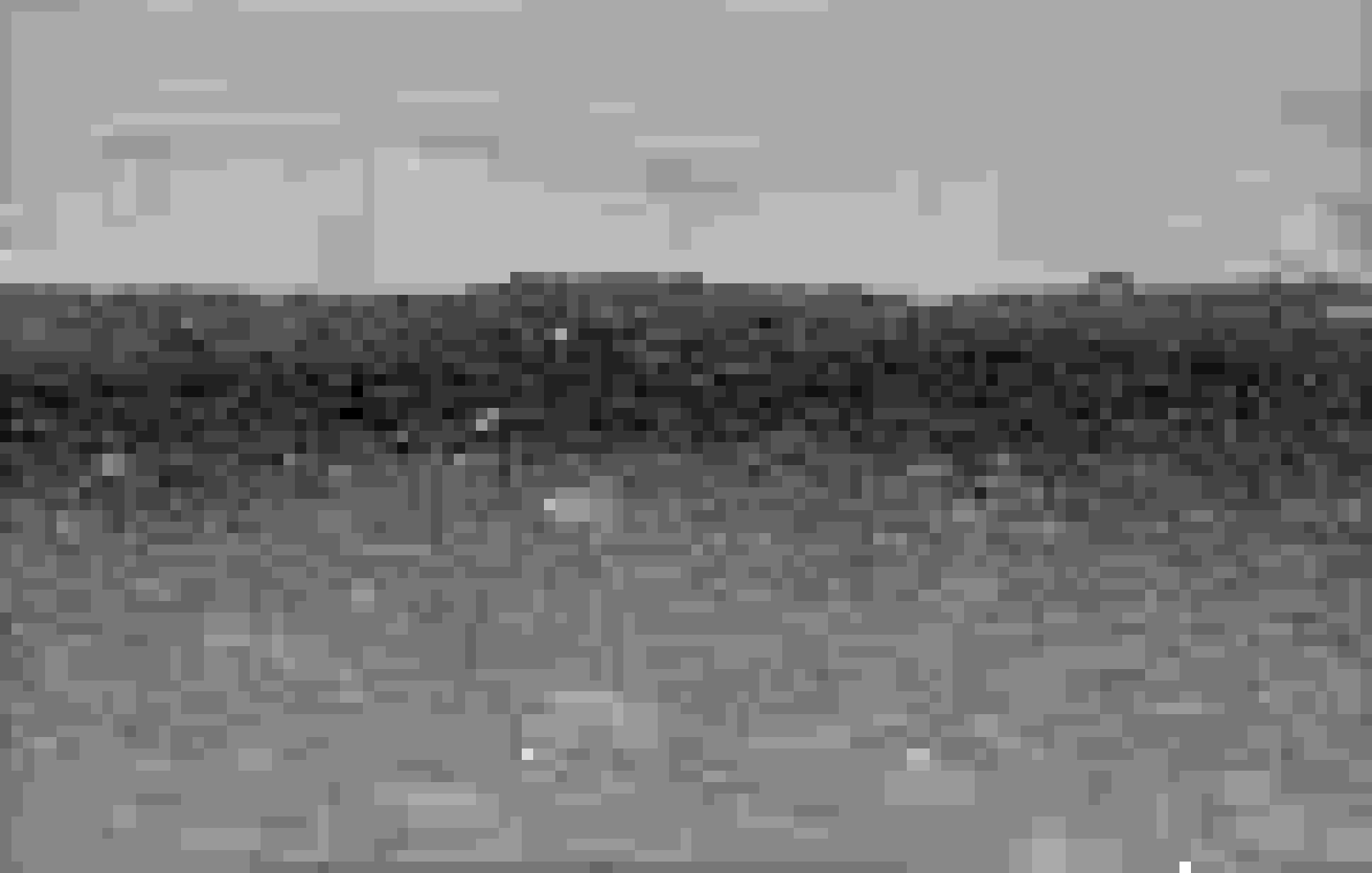}  &
\includegraphics[width=2.5cm]{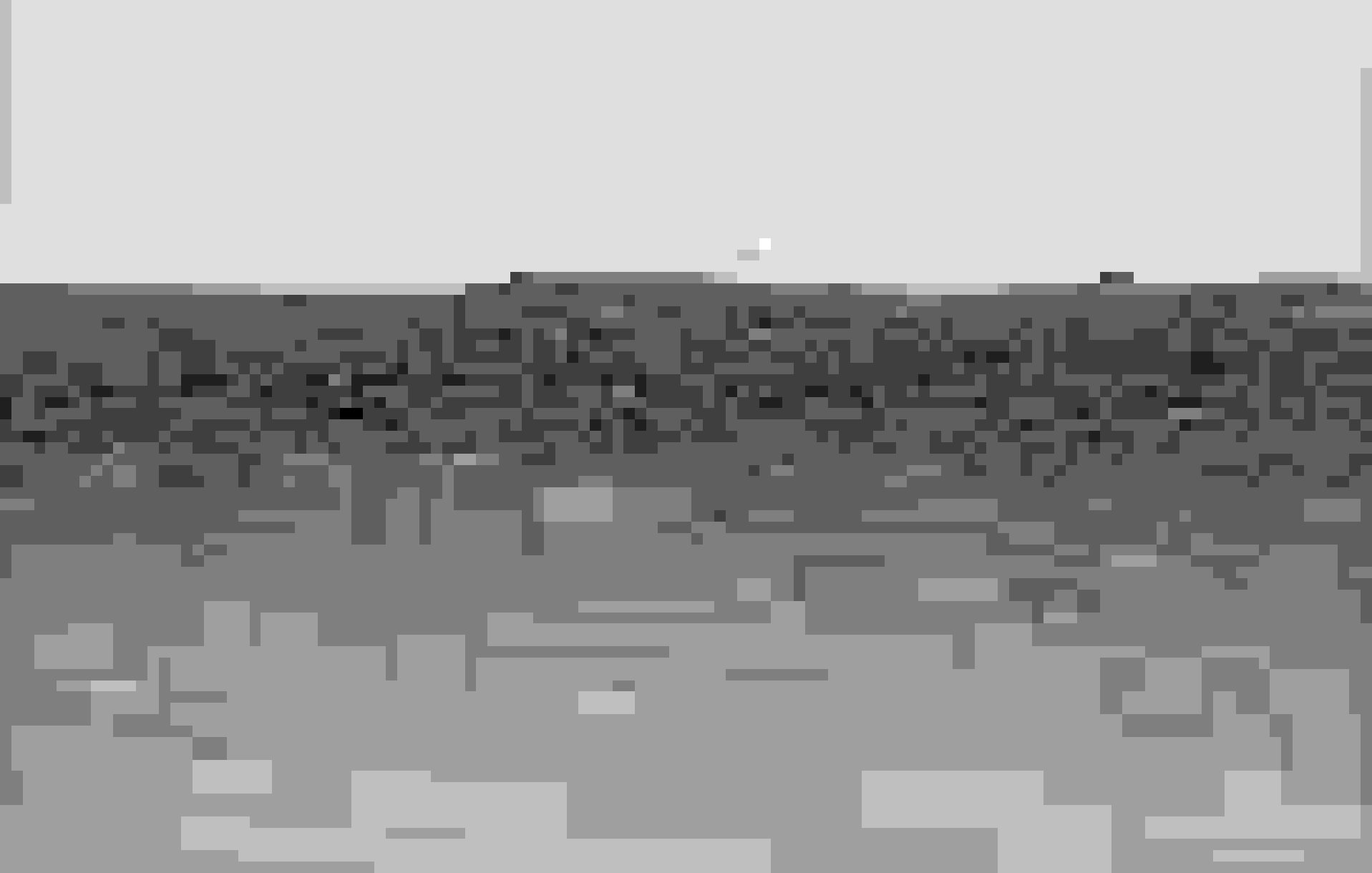}  &
\includegraphics[width=2.5cm]{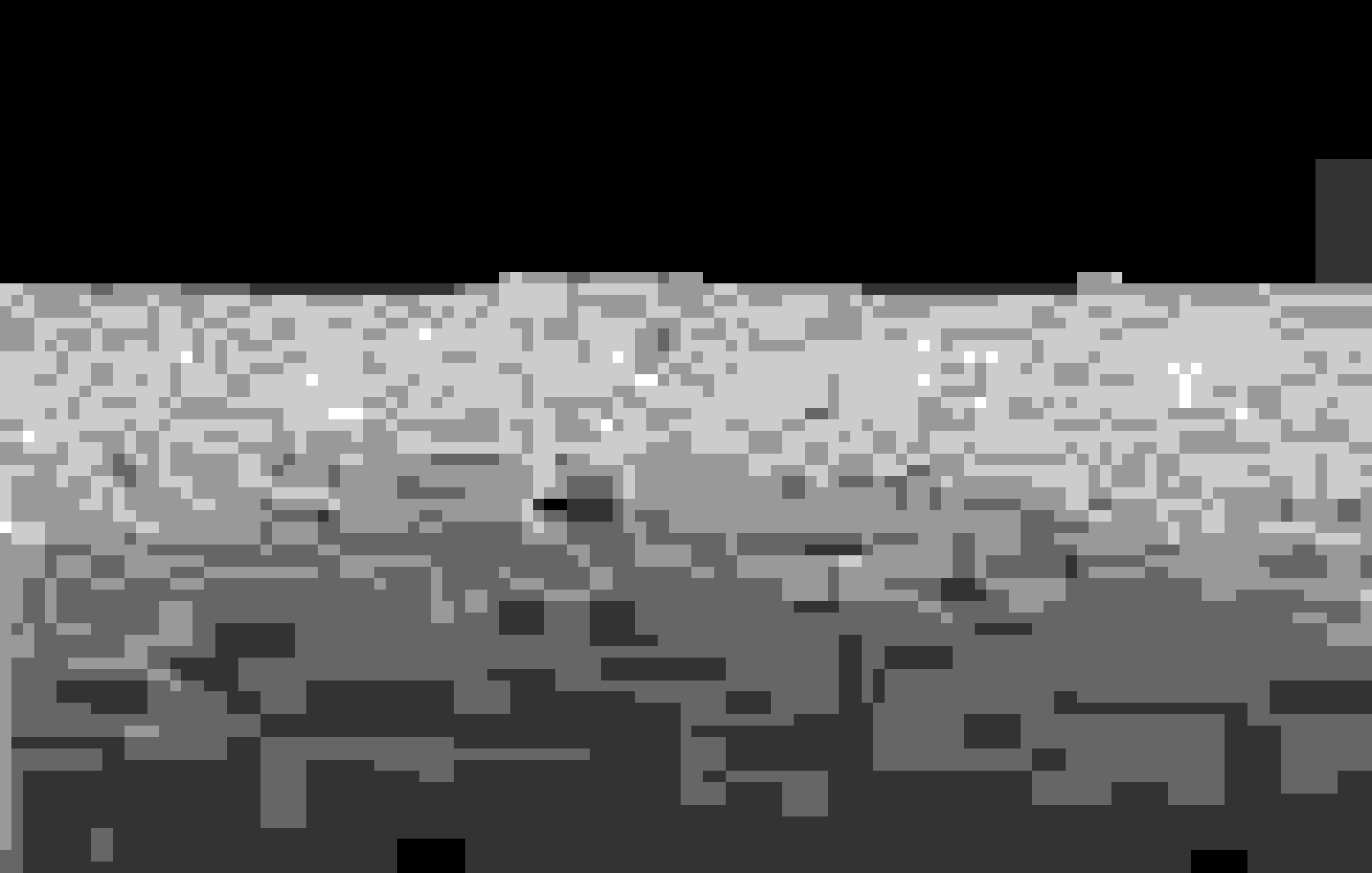}  &
\includegraphics[width=2.5cm]{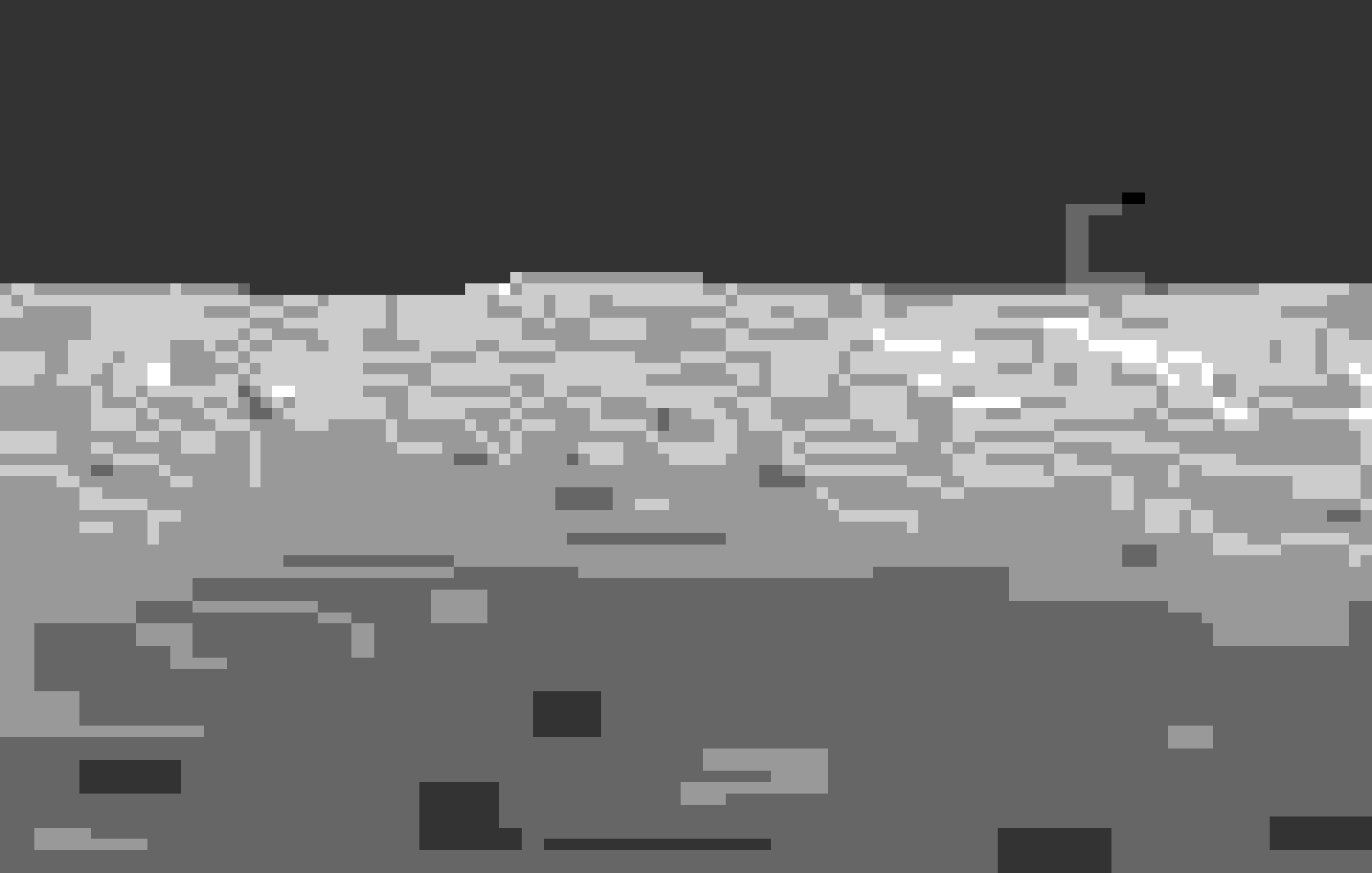}  &
\includegraphics[width=2.5cm]{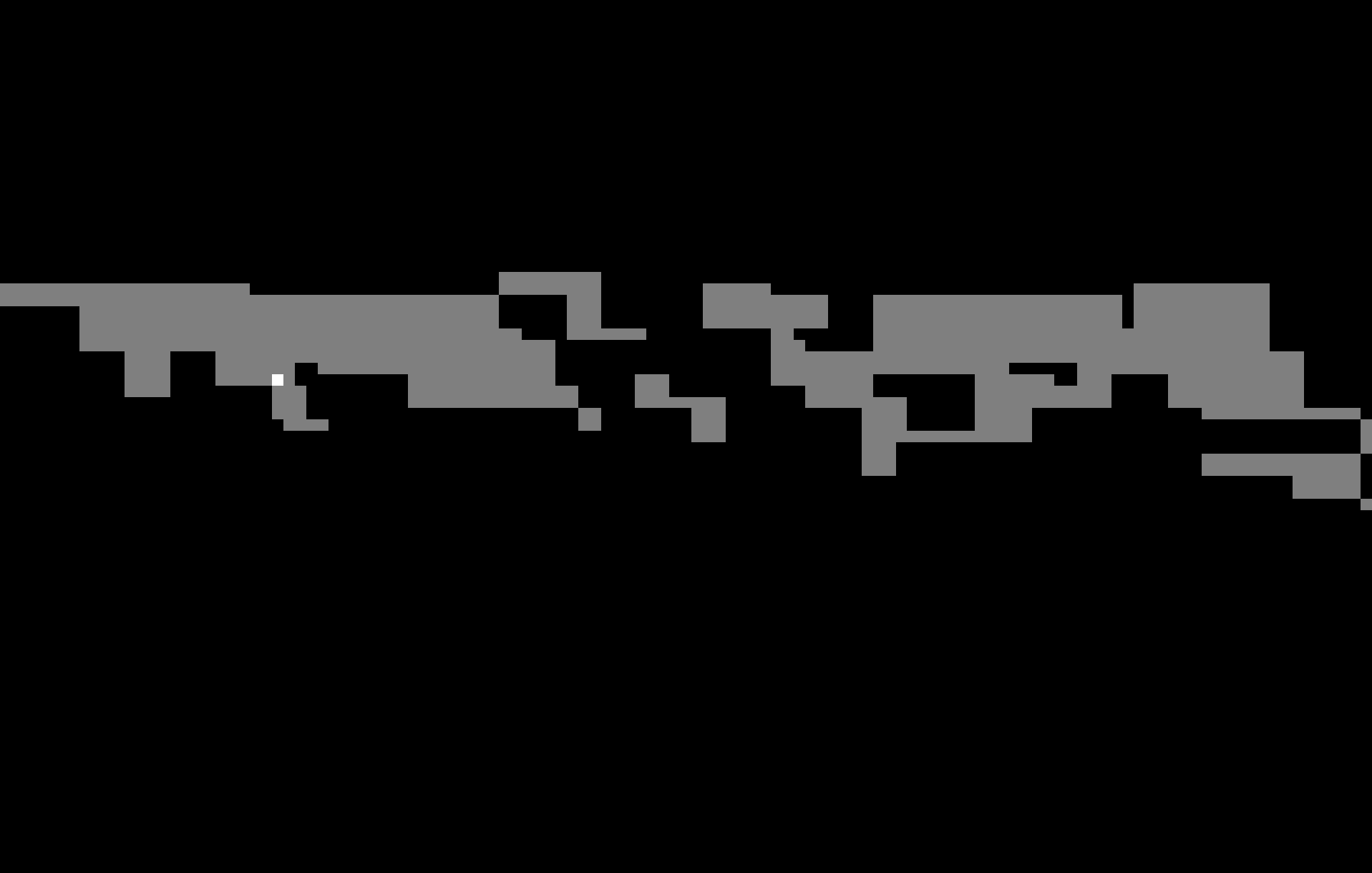}  \\

% roberto-nickson-48063
\raisebox{0.8cm}{\parbox[c]{2cm}{\centering CLIC\\roberto-nickson\\BD-rate -2.88\% \\ $2048\times1365$}} & 
\includegraphics[width=2.5cm]{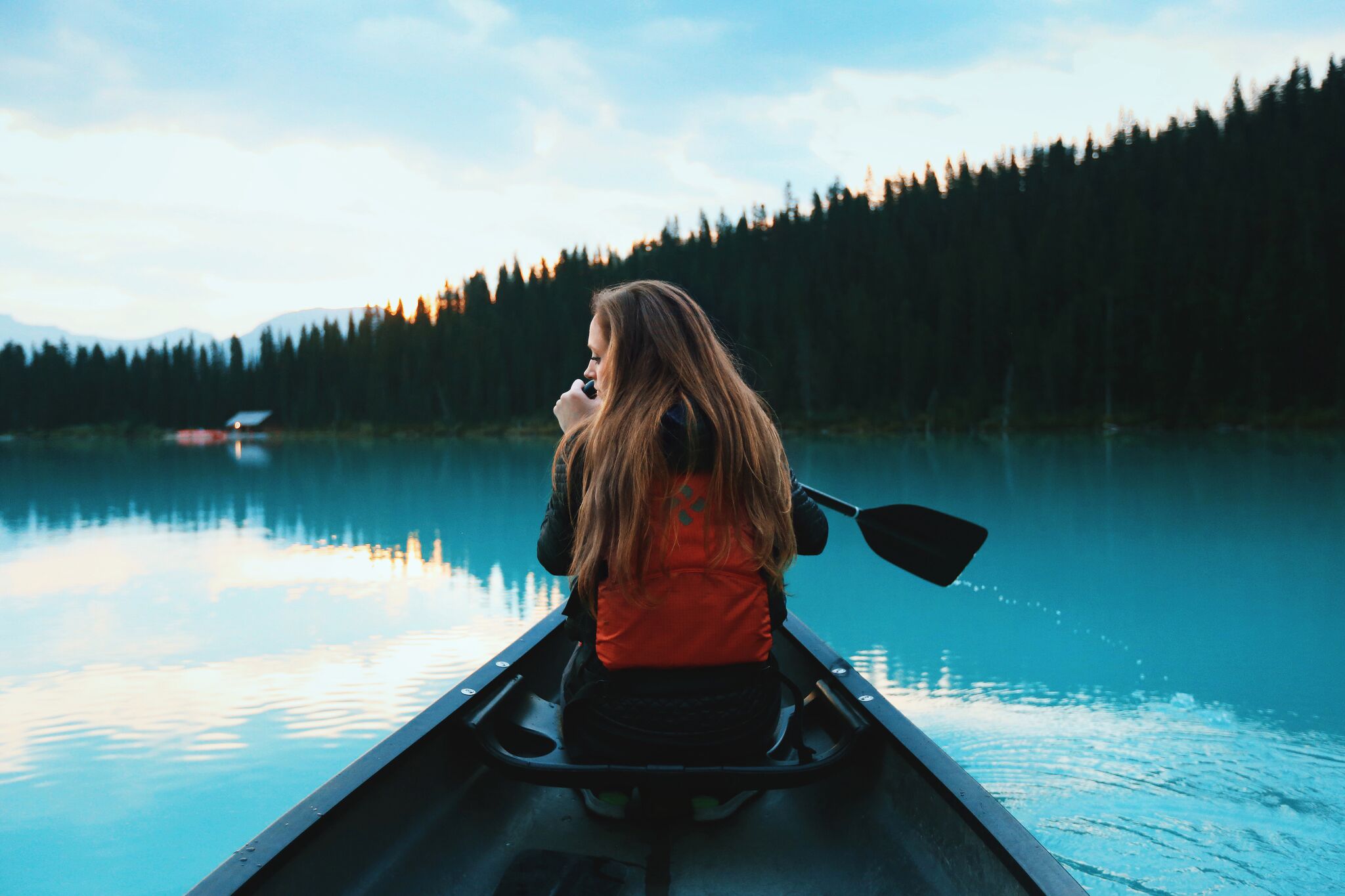} &
\includegraphics[width=2.5cm]{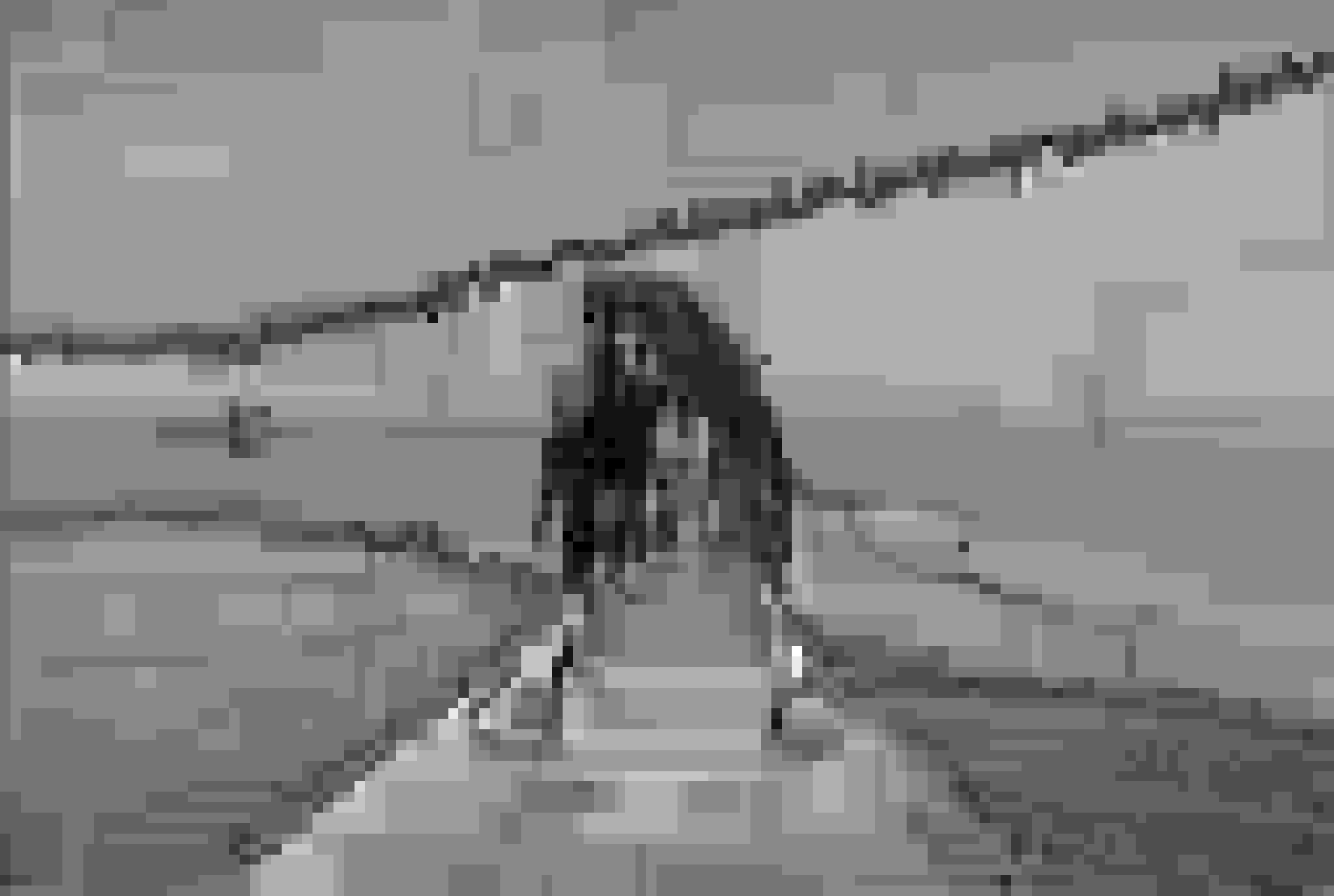}  &
\includegraphics[width=2.5cm]{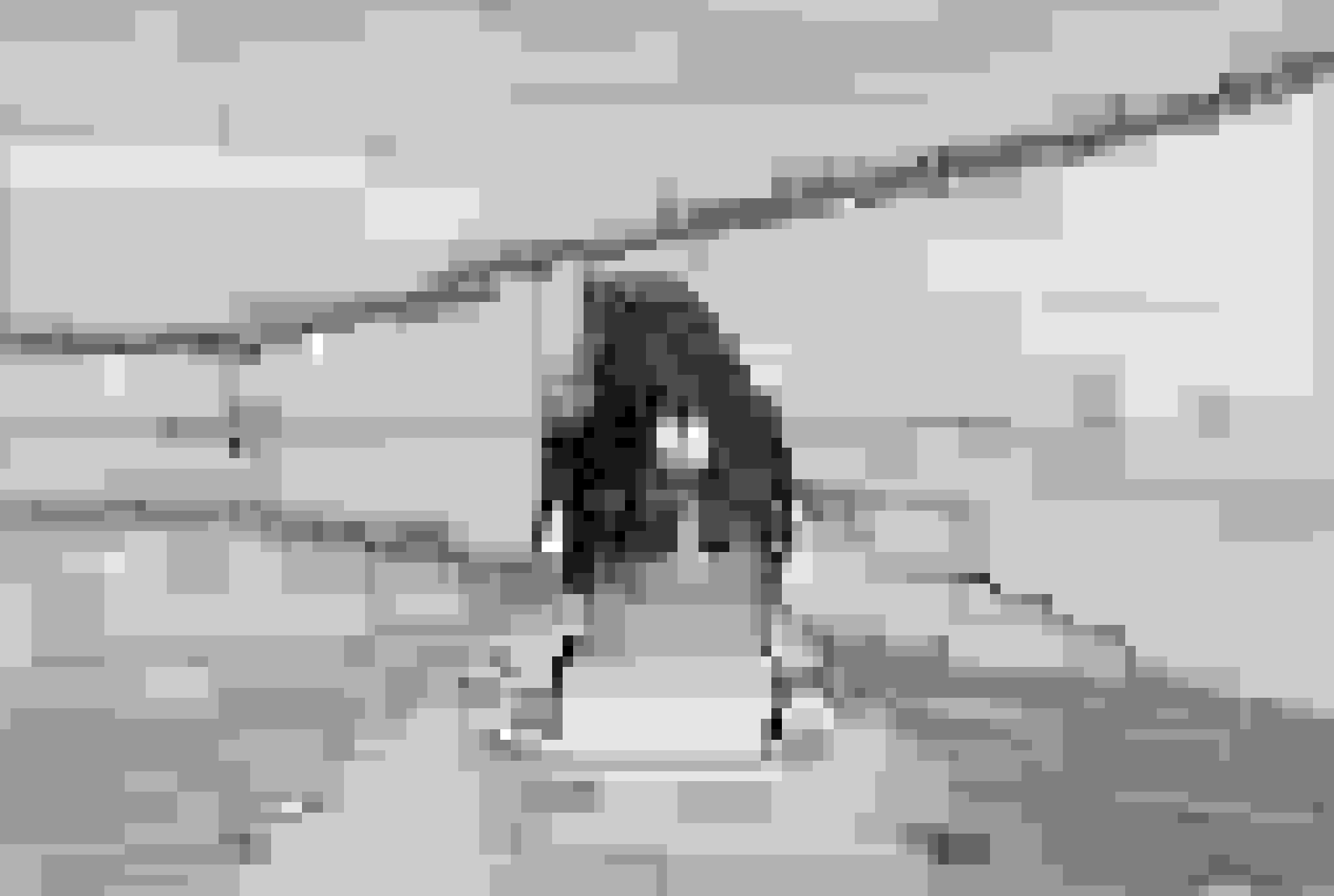}  &
\includegraphics[width=2.5cm]{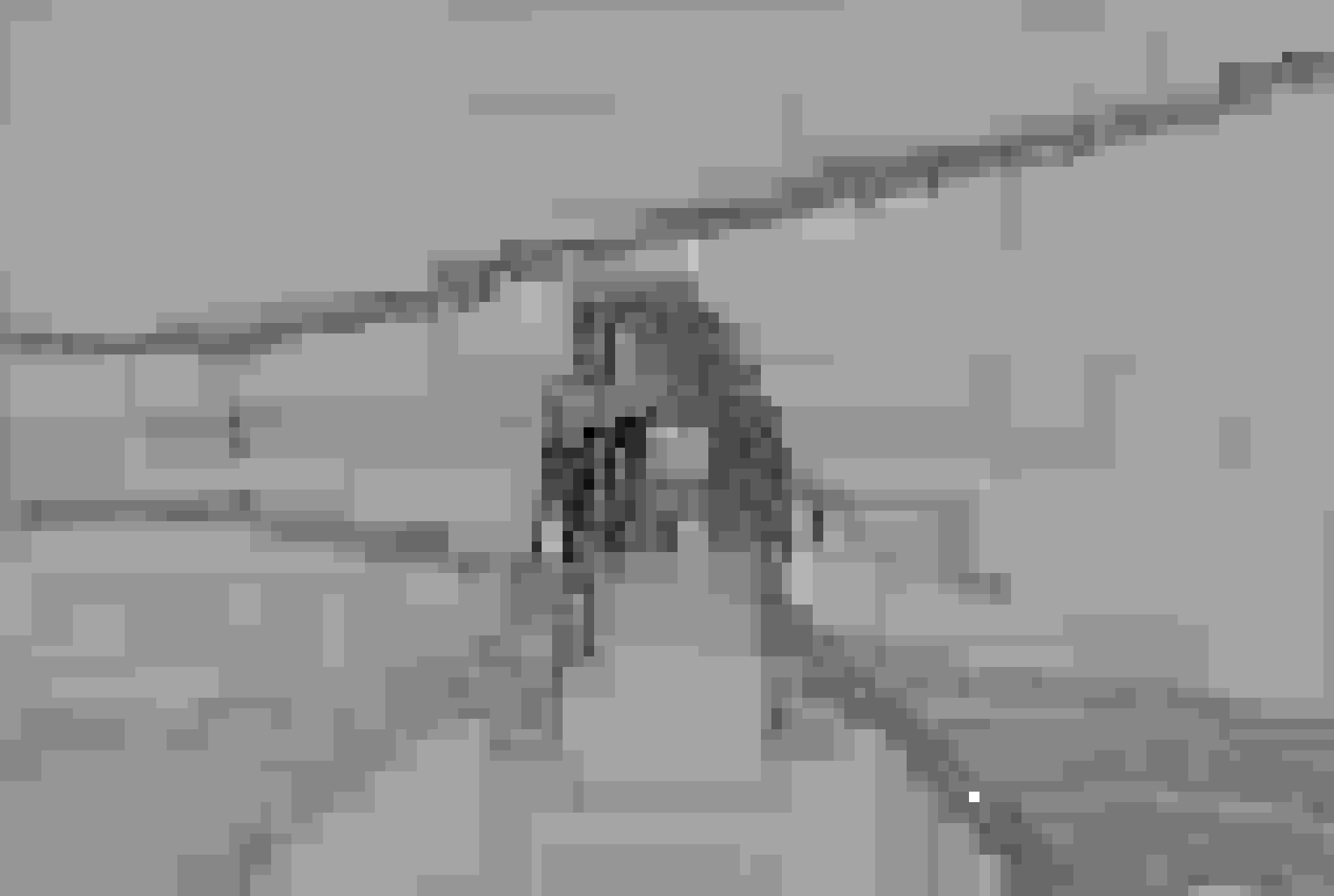}  &
\includegraphics[width=2.5cm]{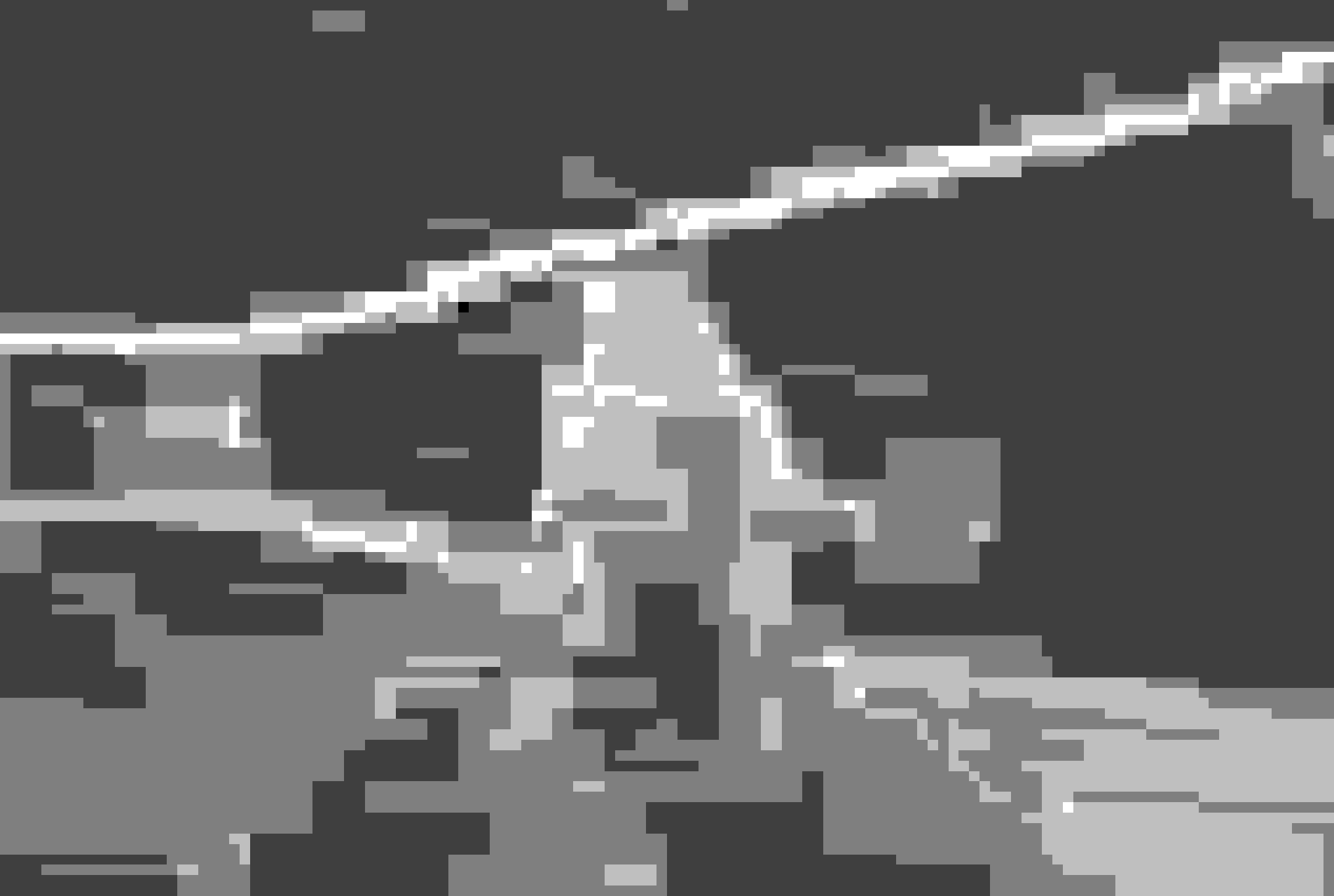}  &
\includegraphics[width=2.5cm]{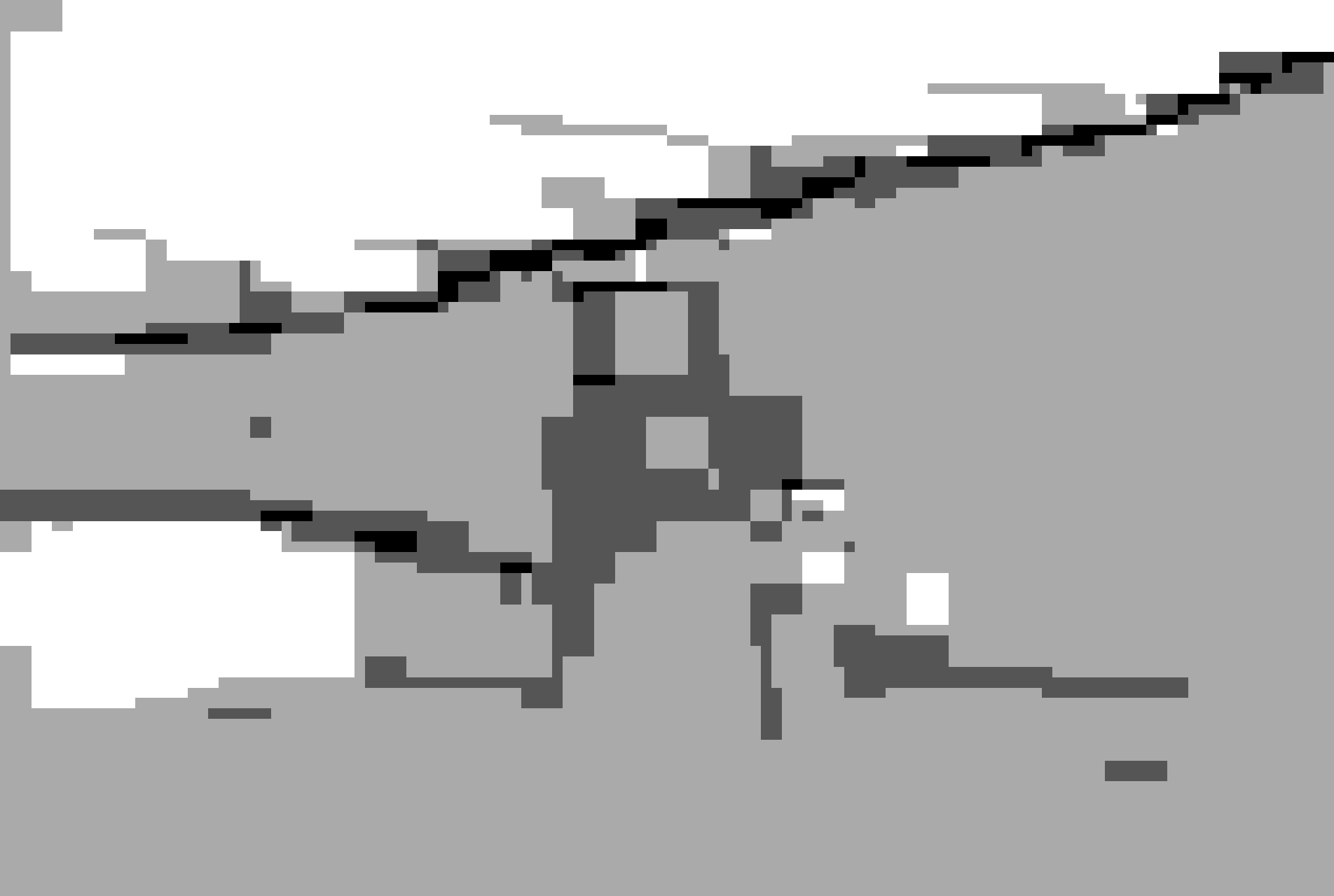}  \\

% zugr-108
\raisebox{0.8cm}{\parbox[c]{2cm}{\centering CLIC\\zugr\\BD-rate -5.44\% \\ $2048\times1365$}} & 
\includegraphics[width=2.5cm]{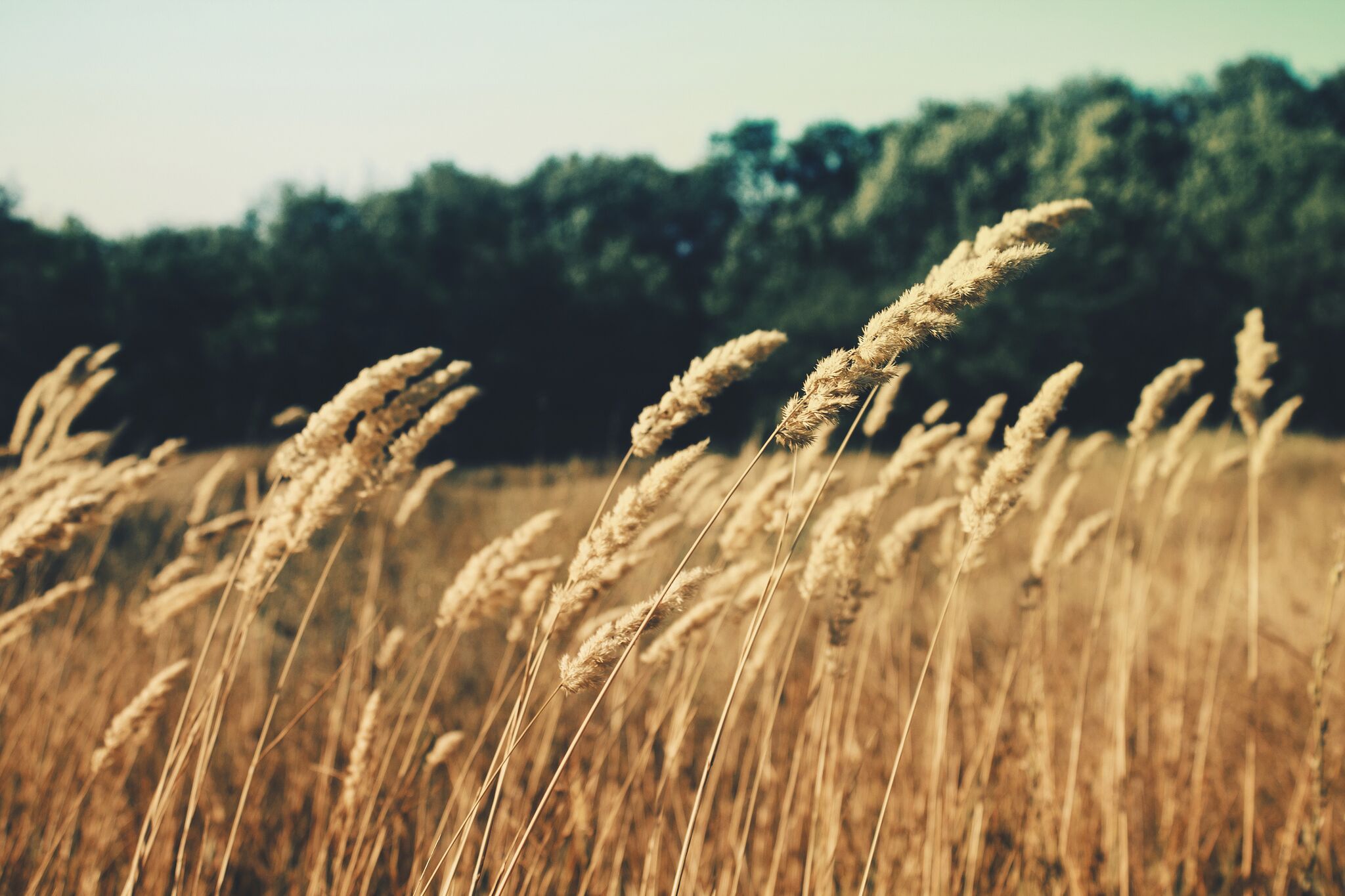} &
\includegraphics[width=2.5cm]{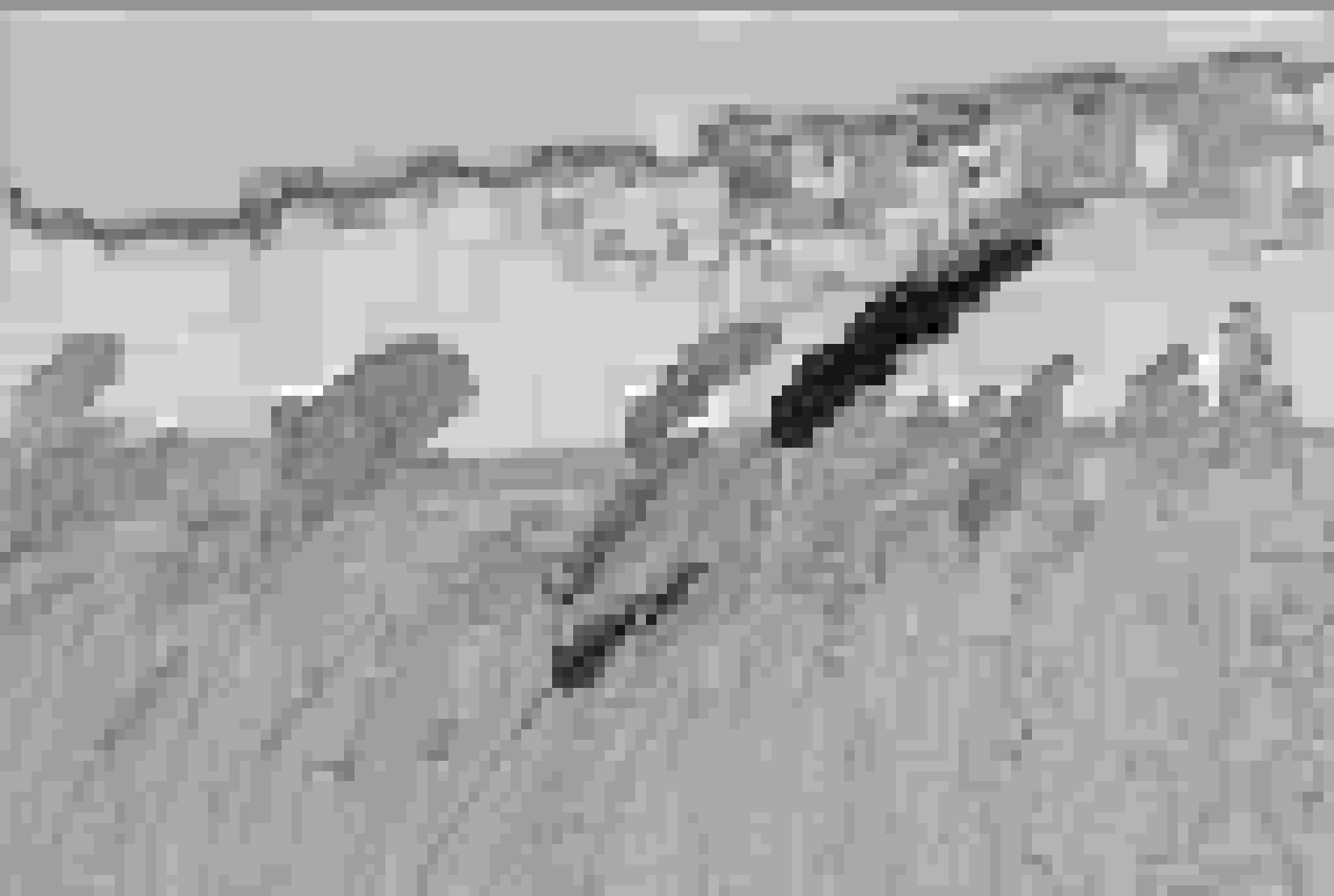}  &
\includegraphics[width=2.5cm]{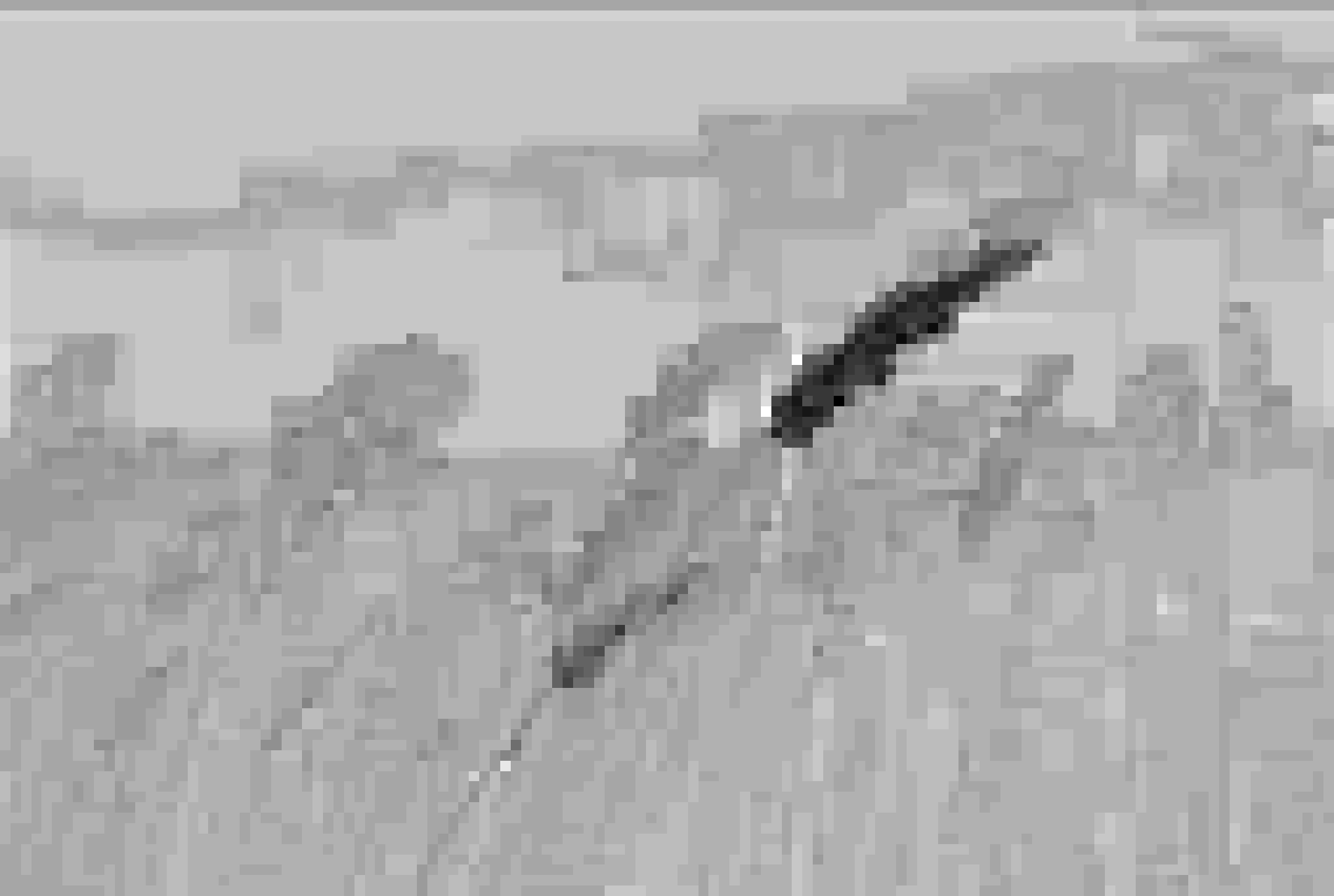}  &
\includegraphics[width=2.5cm]{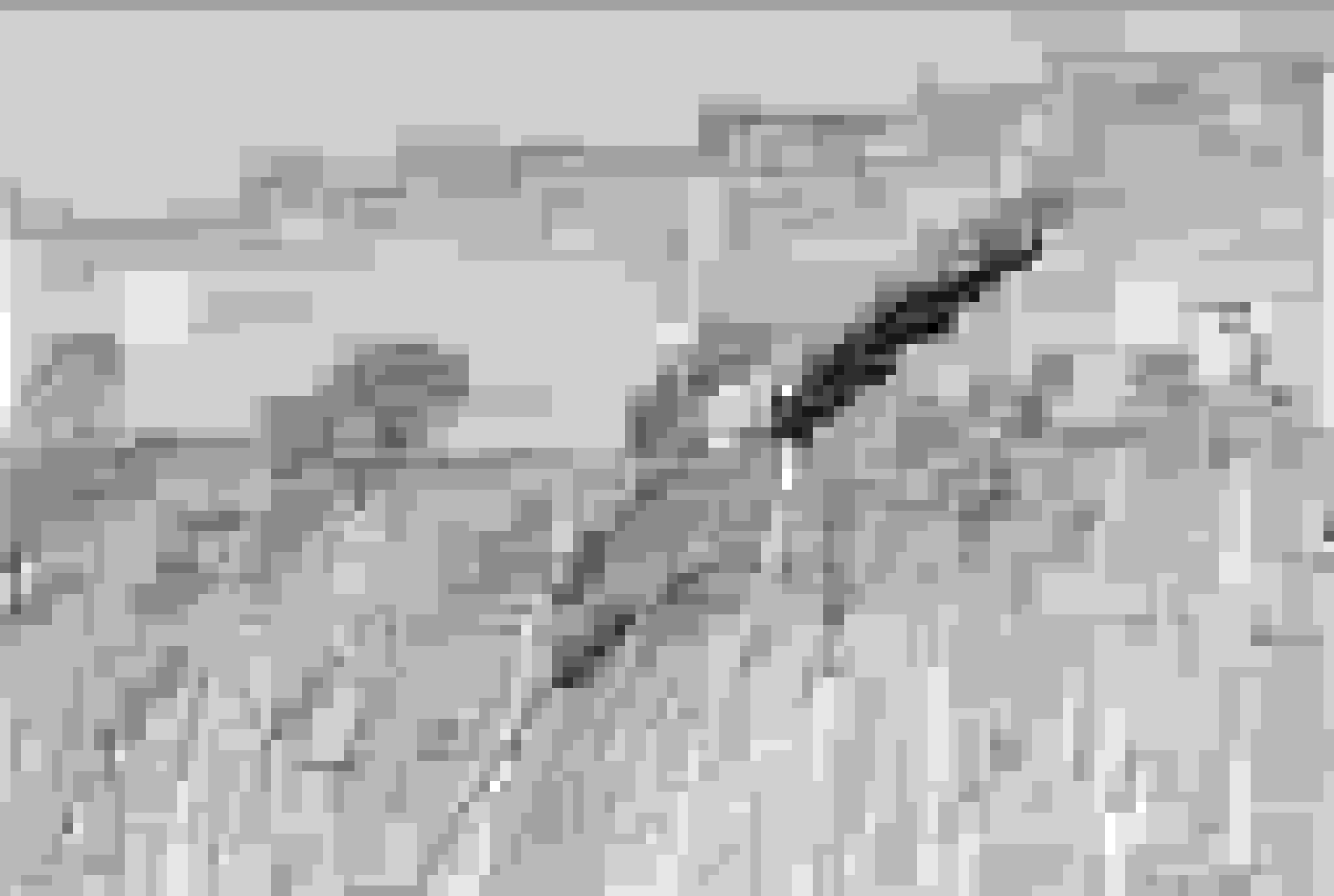}  &
\includegraphics[width=2.5cm]{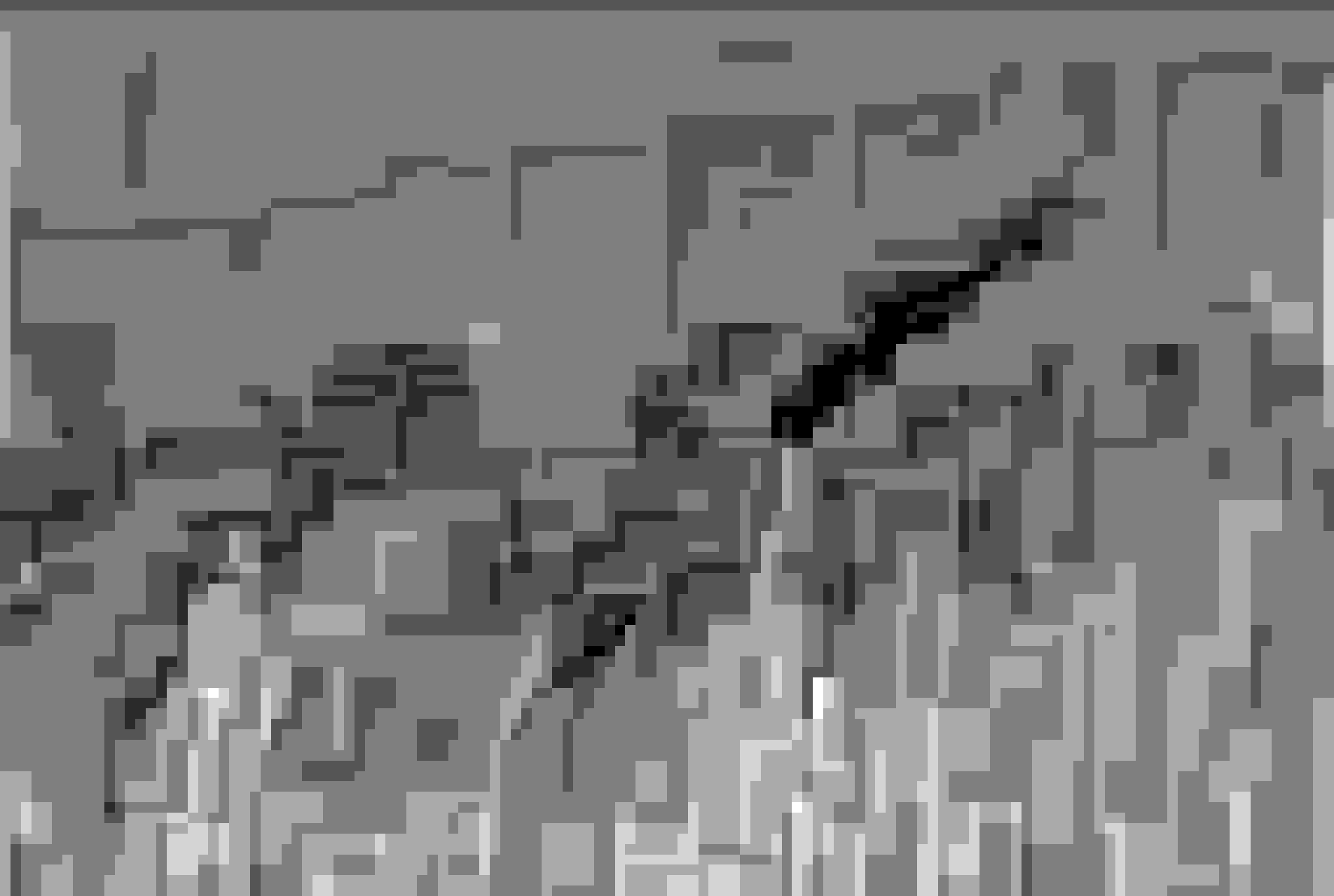}  &
\includegraphics[width=2.5cm]{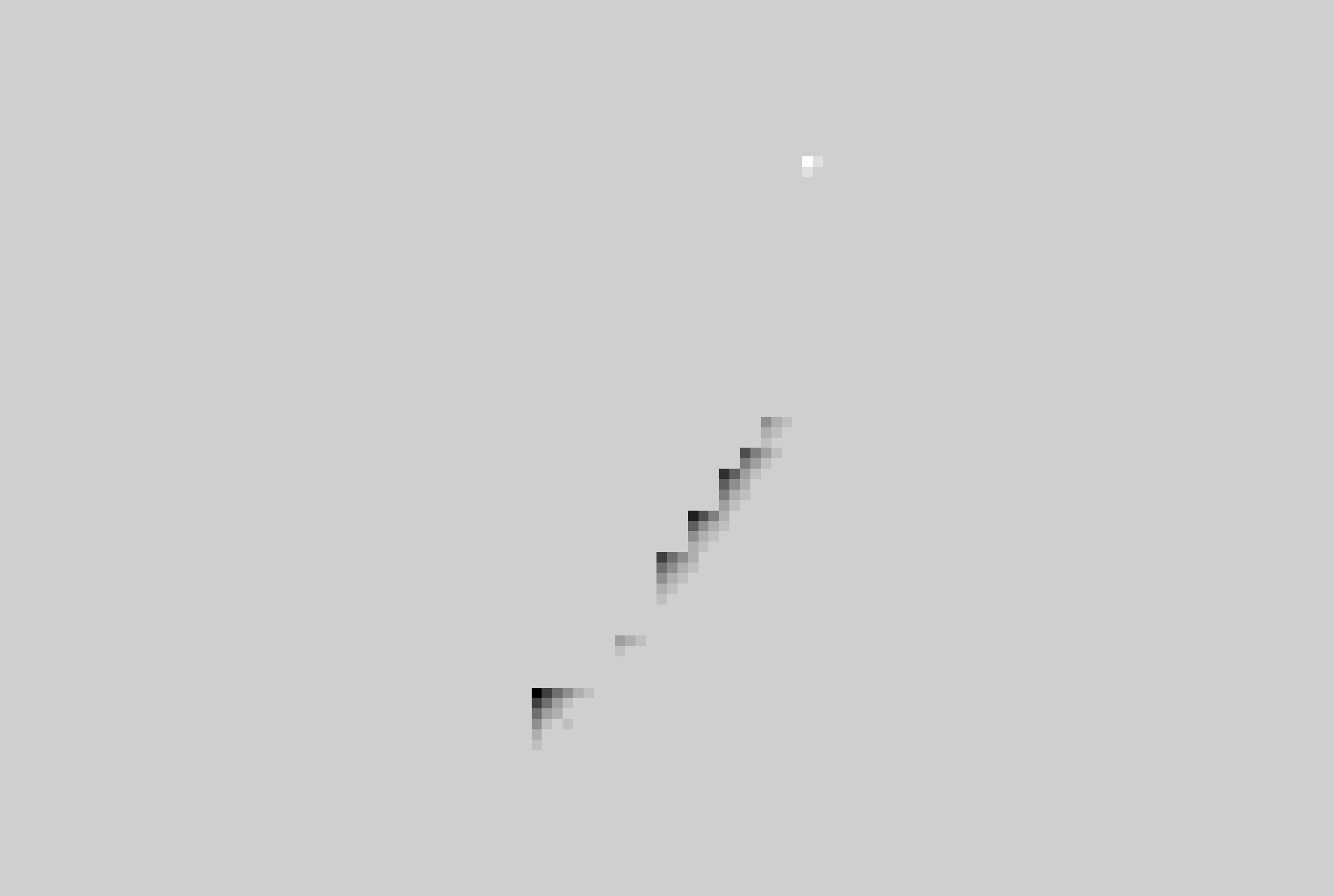}  \\

\end{tabular}
\end{adjustbox}
\caption{Visualization of normalized spatial hyperprior $\bm{\hat{s}}$ maps at different $\lambda$-values. On the left side of each row, the dataset, image name, BD-rate (versus Cool-Chic 4.0) and resolution of the example images is given. Higher $\lambda$-values correspond to lower bitrates.}
\label{fig:example_images}
\end{figure*}

\begin{figure*}[tp]
	\centering
	\includegraphics[width=1\textwidth]{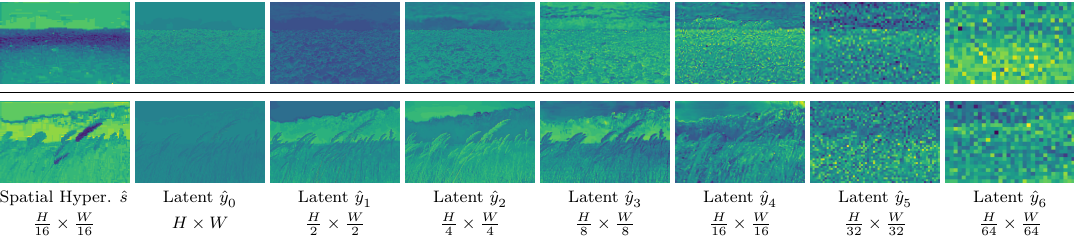}
	\caption{Visualization of the normalized spatial hyperprior maps $\bm{\hat{s}}$ and normalized latents $\bm{\hat{y}}$ for jeremy-cai (top row) and zugr (bottom row) form CLIC 2020 professional validation \cite{clic2020} encoded with $\lambda=0.0001$.}
	\label{fig:spatial_hyperprior_and_latents}
\end{figure*}

Fig.~\ref{fig:example_images} visualizes the spatial hyperprior $\bm{\hat{s}}$ at different $\lambda$-values. The quality of $\bm{\hat{s}}$ degrades with decreasing bitrates (increasing $\lambda$). The images \textit{zugr} and \textit{jeremy-cai} have a shallow depth-of-field and $\bm{\hat{s}}$ appears to map the image sharpness. Only one wheat head in \textit{zugr} is in focus and it is assigned a unique region in $\bm{\hat{s}}$, the other wheat heads are similarly blurry to each-other and are assigned different slightly varying values. The pebbles in \textit{jeremy-cai} get sharper, the further away they are from the camera until they get blurry again. $\bm{\hat{s}}$ at $\lambda\in \{0.0001,0.0004\}$ shows a clear gradient that corresponds to the sharpness of the pebbles with the sharpest pebbles being represented by the darkest color. The body of water in the background is assigned another unique value. Fig.~\ref{fig:spatial_hyperprior_and_latents} shows that this behavior is only visible in $\bm{\hat{s}}$ while the latents $\bm{\hat{y}}$ contain more structural information. For the image \textit{kodim01}, $\bm{\hat{s}}$ appears segmented into unique values for wall and windows/doors. \textit{Kodim15}, \textit{kodim07}, and \textit{roberto-nickson} show similar behavior to this with $\bm{\hat{s}}$ clearly segmenting the image into different regions. In \textit{kodim15}, the face of the girl is even segmented into a bright and a shaded area.

The observed characteristics of $\bm{\hat{s}}$ confirm our assumption that it signals local changes in non-stationary image statistics. Some images have hard boundaries between regions of similar image statistics leading to $\bm{\hat{s}}$ that appear segmented. Other images have gradual changes in image statistics leading to the gradual changes in $\bm{\hat{s}}$. The better performance of LANCE on CLIC compared to Kodak shows that it performs better on high-resolution images. Summarizing, LANCE yields the highest performance gains for images characterized by high variations in regional image statistics, such as those featuring distinct foreground-background separation, shallow depth-of-field, or significant variations in local illumination. The spatial complexity \cite{menon2022vca} of the images is uncorrelated ($R^2=0.001$) to the BD-rate and there is only an weak correlation ($R^2=0.15$) between the compressed size of $\bm{\hat{s}}$ and the BD-rate.

\begin{table}[h!]
    \centering
    \caption{Decoder complexity of our method LANCE and Cool-Chic 4.0 in MAC/pixel for three different operation points. }
    \label{tab:decoder_complexety_values}
    \setlength{\tabcolsep}{2pt}
    \begin{tabular}{cc|ccc|cc}
    \toprule
    &&\multicolumn{3}{c|}{Exist in Cool-Chic 4.0}&\multicolumn{2}{c}{LANCE} \\
    Method&\shortstack{Operation\\point} & \shortstack{Context\\ $C_\xi$} & \shortstack{Upsampler\\  $f_\upsilon$} & \shortstack{Synthesis\\ $f_\theta$} & \shortstack{Context\\ $C_\phi$ } & Resampler \\
    \midrule
    \multirow{3}{*}{LANCE}&HOP & 768.15 & 66.67 & 642.0 & 0.06 & 6.38 \\
    &MOP & 768.15 & 66.67 & 322.0 & 0.06 & 6.38 \\
    &LOP & 213.37 & 66.67 & 322.0 & 0.06 & 6.38 \\
    \midrule
    \multirow{3}{*}{\shortstack{Cool-Chic\\4.0}}&HOP & 725.29 & 66.67 & 642.0 & - & - \\
    &MOP & 725.29 & 66.67 & 322.0 & - & - \\
    &LOP & 191.99 & 66.67 & 322.0 & - & - \\
    \bottomrule
    \end{tabular}
\end{table}
Table~\ref{tab:decoder_complexety_values} shows a per-module breakdown of the decoder complexity. The complexity increase of LANCE is mostly caused by the two additional inputs of the context model $C_\xi$. The complexity of the spatial prior context model $C_\phi$ is negligible. The MED-predictor does not perform any multiplications and thus has a complexity of 0 MAC/pixel. The complexity of the bicubic resampler is orders of magnitude higher than $C_\phi$, but still low compared to the complexity increase of $C_\xi$. The Upsampler and Synthesis of Cool-Chic and LANCE have identical architectures which consequently leads to identical complexities.

Table~\ref{tab:runtimes} compares encoder and decoder execution times evaluated on the Kodak dataset using an Intel Core i9-9900K CPU @ 5.00 GHz. Decoding times are CPU only using decoder implementations written in C++. Encoding times were measured using an Nvidia RTX A6000 GPU. On average, LANCE has a 1.3 times higher encoding time and a 1.1 times higher decoding time compared to Cool-Chic 4.0. 

Table~\ref{tab:decoder_complexety_runtimes} shows that the increase in decoding time relative to Cool-Chic 4.0 is mainly caused by the bicubic resampler followed by the more complex context model $C_\xi$.

\subsection{Bitstream Breakdown}

Table~\ref{tab:byte_percentage} shows the relative signaling cost for the different parts of our codec. Similar to Cool-Chic, the relative signaling cost for the neural network weights decreases with increasing bitrate. This is to be expected, since the neural network architecture is identical for all bitrates leaving the quantization of parameters as the only variable while more and more information has to be signaled in the latent $\bm{\hat{y}}$. Even though the absolute rate of the spatial hyperprior $\bm{\hat{s}}$ increases with raising bitrates (from 194 Bytes at $\lambda=0.02$ to 1212 Bytes at $\lambda=0.0001$ on CLIC), the relative portion of its rate decreases with increasing bitrate. In other words: the signaling cost of $\bm{\hat{y}}$ grows faster than the cost of $\bm{\hat{s}}$ with raising bitrates.

\begin{table}[h]
    \centering
    \caption{Average decoder and encoder execution times for LANCE and Cool-Chic 4.0 for three different operation points.}
    \label{tab:runtimes}
    \setlength{\tabcolsep}{4pt}
    \begin{tabular}{c|ccc|ccc}
    \toprule
    &\multicolumn{3}{c|}{Decoder runtimes [ms]}&\multicolumn{3}{c}{Encoder runtimes [min]} \\
    & HOP & MOP & LOP & HOP & MOP & LOP \\
    \midrule
    LANCE& 49.45 & 47.29 & 42.74 & 33.65 & 31.19 & 29.12 \\
    \midrule
    Cool-Chic& 46.13 & 42.36 & 38.10 & 26.45 & 23.88 & 22.13 \\
    \bottomrule
    \end{tabular}
\end{table} 
\begin{table}[h!]
    \centering
    \caption{Decoder execution times of our method LANCE and Cool-Chic 4.0 in ms for three different operation points. Note that execution times for $C_\xi$ and $C_\phi$ include entropy decoding.}
    \label{tab:decoder_complexety_runtimes}
    \setlength{\tabcolsep}{2pt}
    \begin{tabular}{cc|ccc|cc}
    \toprule
    &&\multicolumn{3}{c|}{Exist in Cool-Chic 4.0}&\multicolumn{2}{c}{LANCE} \\
    Method&\shortstack{Operation\\point} & \shortstack{Context\\ $C_\xi$} & \shortstack{Upsampler\\  $f_\upsilon$} & \shortstack{Synthesis\\ $f_\theta$} & \shortstack{Context\\ $C_\phi$ } & Resampler \\
    \midrule
    \multirow{3}{*}{LANCE}&HOP & 22.45 & 3.80 & 14.23 & 0.07 & 2.85 \\
    &MOP & 22.96 & 3.81 & 11.59 & 0.07 & 2.94 \\
    &LOP & 18.31 & 3.84 & 11.61 & 0.08 & 2.99 \\
    \midrule
    \multirow{3}{*}{\shortstack{Cool-Chic\\4.0}}&HOP & 20.45 & 3.84 & 14.22 & - & - \\
    &MOP & 20.51 & 3.82 & 11.60 & - & - \\
    &LOP & 16.63 & 3.91 & 11.67 & - & - \\
    \bottomrule
    \end{tabular}
\end{table} 
\begin{table}[h!]
    \centering
    \caption{Average relative bitstream composition for the HOP complexity operation point of LANCE evaluated on Kodak \cite{kodak} and CLIC 2020 professional \cite{clic2020}. The difference between the row sum and 100\% is due to the header overhead.}
    \setlength{\tabcolsep}{3pt}
    \begin{tabular}{cc|cccc|cc}
\toprule
\multicolumn{8}{c}{Kodak} \\
\midrule
&&\multicolumn{4}{c|}{Exist in Cool-Chic 4.0}&\multicolumn{2}{c}{LANCE} \\
$\lambda$ & Size [KB] & $\bm{\hat{y}}$ [\%] & $C_\xi$ [\%] & $f_\upsilon$ [\%] & $f_\theta$ [\%] & $C_\phi$ [\%] & $\bm{\hat{s}}$ [\%] \\
\midrule
0.0001 & 67.85 & 96.99 & 0.97 & 0.10 & 1.36 & 0.03 & 0.42 \\
0.0004 & 38.14 & 95.02 & 1.65 & 0.16 & 2.25 & 0.05 & 0.65 \\
0.001 & 24.48 & 92.62 & 2.52 & 0.24 & 3.35 & 0.07 & 0.85 \\
0.004 & 11.02 & 85.13 & 5.40 & 0.52 & 6.76 & 0.16 & 1.28 \\
0.02 & 4.13 & 64.79 & 14.20 & 1.35 & 15.95 & 0.37 & 1.37 \\
\midrule
\multicolumn{8}{c}{CLIC} \\
\midrule
&&\multicolumn{4}{c|}{Exist in Cool-Chic 4.0}&\multicolumn{2}{c}{LANCE} \\
$\lambda$ & Size [KB] & $\bm{\hat{y}}$ [\%] & $C_\xi$ [\%] & $f_\upsilon$ [\%] & $f_\theta$ [\%] & $C_\phi$ [\%] & $\bm{\hat{s}}$ [\%] \\
\midrule
0.0001 & 266.21 & 98.86 & 0.25 & 0.03 & 0.38 & 0.01 & 0.44 \\
0.0004 & 133.44 & 98.02 & 0.50 & 0.05 & 0.70 & 0.01 & 0.64 \\
0.001 & 82.89 & 97.09 & 0.79 & 0.08 & 1.07 & 0.02 & 0.83 \\
0.004 & 36.82 & 94.30 & 1.71 & 0.17 & 2.22 & 0.05 & 1.31 \\
0.02 & 12.57 & 86.79 & 4.75 & 0.47 & 5.69 & 0.13 & 1.51 \\
\bottomrule
\end{tabular}
    \label{tab:byte_percentage}
\end{table}

The latent $\bm{\hat{y}}$ contributes more to the overall bitrate on CLIC compared to Kodak. This makes sense, since the CLIC images have higher resolutions, increasing the latent dimension and costs while the number of neural network parameters stays constant. The spatial hyperprior $\bm{\hat{s}}$ behaves in a similar way for the same reason, but especially at low bitrates, the relative rate difference between Kodak and CLIC is higher for $\bm{\hat{y}}$ than for $\bm{\hat{s}}$: $R(\bm{\hat{s}})/(R(\bm{\hat{s}}) + R(\bm{\hat{y}}))$ is 2.1 \% and 0.44 \% for Kodak, and 1.7 \% and 0.45 \% for CLIC at $\lambda=0.02$ and $\lambda=0.0001$, respectively.

We can conclude from this and the only weak correlation between $R(\bm{\hat{s}})$ and BD-rate, that higher $R(\bm{\hat{s}})$ do not necessarily indicate higher rate distortion performance. This makes sense, since $R(\bm{\hat{s}})$ and the rate distortion performance gain caused by $\bm{\hat{s}}$ depend on different image characteristics: $R(\bm{\hat{s}})$ depends on the conditional cross-entropy of $\bm{\hat{s}}$ and by that (under the assumption that $\bm{\hat{s}}$ signals local changes in image statistics) on how well $C_\phi$ and the MED-predictor can model the dependencies between the local changes in image statistics. The rate-distortion performance gain of LANCE on the other hand depends on how varied the image statistics are regionally. The spatial hyperprior can have easy to predict spatial dependencies for some images while it still gives good gains. The opposite case (and everything in between) can occur as well, where the spatial hyperprior does not give that high rate-distortion gains and is also hard to predict. All this indicates, that the performance of LANCE is mostly content dependent.

\subsection{Ablation Study}
The ablation results in Table \ref{tab:ablation} demonstrate the individual and joint contributions of the LANCE sub-modules alongside their respective impacts on encoding/decoding run times and computational complexity. The layer index $\bar{l}$ alone provides a modest BD-rate improvement of -0.35\%. The spatial hyperprior $\bm{\hat{s}}$ is the primary driver of performance. When used in isolation without $\bar{l}$, the spatial hyperprior achieves a significantly higher gain of -0.84\%, confirming its ability in capturing local image statistics.

The choice of resampling method for the hyperprior also impacts efficiency. Replacing the bicubic resampler with area resampling (average of input pixel values covered by the output pixel) reduces the BD-rate gain from -1.40\% to -1.14\%, suggesting that bicubic interpolation more effectively maps the low-dimension hyperprior to the latent space.

Furthermore, the combination of the static MED-predictor $\tilde{s}_{\text{MED}}$ with the learned context model $C_\phi$ is a good context estimation architecture for $\bm{\hat{s}}$. Utilizing $C_\phi$ or $\tilde{s}_{\text{MED}}$ individually results in BD-rate gains of -1.30\% and -1.24\%, respectively. The full LANCE configuration, combining the layer index with the complete spatial hyperprior pipeline, yields the best BD-rate of -1.40\%.

In the test with only $\bm{\tilde{s}}^{(\text{MED})}$ activated, $\sigma^{(\phi)}_{ij}$ is estimated by learning and signaling a single parameter for it for the whole image. We tested learning different $\sigma^{(\phi)}_{ij}$ parameters for different contexts split based on the local gradient as in \cite{weinberger2000loco}, too. However, this did not lead to any gains compared to learning a only single value.

The resampler is the primary driver of computational overhead; switching from the bicubic resampler to area resampling reduces the relative encoding time from 1.27 to 1.05 and decoding time from 1.07 to 1.02. The layer index $\bar{l}$ has only a slight impact on the decoding time and no significant impact on the encoding time. The context estimation has no relevant impact on encoding or decoding time.

\begin{table}[h!]
    \centering
    \caption{BD-rates and relative encoding and decoding run times of LANCE using the HOP complexity operation point tested with sub-modules turned on and off vs. Cool-Chic 4.0 evaluated on Kodak \cite{kodak}. }
    \label{tab:ablation}
    \setlength{\tabcolsep}{2pt}
    \begin{tabular}{cccccccc}
        \toprule
        \multirow{2}*{\shortstack{Layer\\Index\\$\bar{l}$}} & 
        \multirow{2}*{\shortstack{Spatial\\Hyper.\\$\bm{\hat{s}}$}} & 
        \multicolumn{2}{c}{Hyperprior Setup} & 
        BD-rate & 
        \multicolumn{2}{c}{Rel. Time} & 
        Dec. Comp. \\
        \cmidrule(lr){3-4} \cmidrule(lr){6-7}
         & & Resampler & Context Est. & [\%] & Enc. & Dec. &  [MAC/px]\\[3pt] % Adds 3pt padding below the text
        \midrule
        \xmark & \xmark  & -&  - & \phantom{-}0.00 &1.00&1.00&1433.96\\
        \cmark & \xmark  & -&  - & -0.35 &1.02&1.01&1455.39\\
        \xmark& \cmark  & bicubic & $\text{MED}+C_\phi$  & -0.84 &1.27&1.06&1461.83\\
        \cmark & \cmark  & area & $\text{MED}+C_\phi$  & -1.14 &1.05&1.02&1476.88\\
        \cmark & \cmark  & bicubic & $\text{MED}$  & -1.24 &1.27&1.07&1483.20\\
        \cmark & \cmark  & bicubic & $C_\phi$ & -1.30\ &1.27&1.07&1483.26\\
        \cmark & \cmark  & bicubic & $\text{MED}+C_\phi$ & \textbf{-1.40} &1.27&1.07&1483.26\\
        \bottomrule
    \end{tabular}
\end{table}

\section{Conclusion}
\label{conclusion}
In this work, we presented LANCE, a method to enhance the context modeling within overfitted image codecs. By introducing a spatial hyperprior that conditions the autoregressive context model, we enabled the codec to handle non-stationary local image statistics. Our hybrid predictive coding approach to signal the spatial hyperprior, utilizing both a static MED-predictor and a small learned context model, ensures that this added expressivity is not negated by the signaling overhead and computational cost.

Experimental results confirmed that LANCE consistently outperforms the Cool-Chic 4.0 baseline by 1.40\% to 2.99\% across all operation points, offering better BD-rates at identical or lower complexities. The ablation study identified the spatial hyperprior as the primary driver of our compression gains.

\section{Acknowledgments}
The authors thank the co-first authors of \cite{MoRIC2025}, Gen Li and Haotian Wu, for helpful discussions regarding the results reported in their work.

%\bibliographystyle{IEEEtran}
%\bibliography{refs}

\newpage

%\bf{If you include a photo:}\vspace{-33pt}
\begin{IEEEbiography}[{\includegraphics[width=1in,height=1.25in,clip,keepaspectratio]{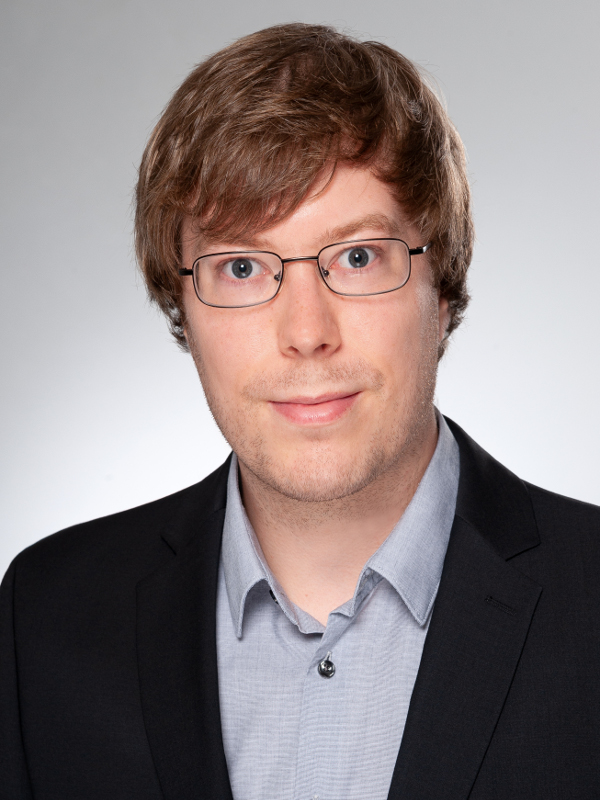}}]{Martin Benjak} received the B.Sc. degree in electrical engineering from the University of Applied Sciences of Osnabrück, in 2017, and the M.Sc. degree in electrical engineering from Leibniz Universität Hannover, in 2019, where he is currently pursuing the doctorate degree with the Institut für Informationsverarbeitung. His research interests include learning-based image/video coding and video coding for machines.
\end{IEEEbiography}
\begin{IEEEbiography}[{\includegraphics[width=1in,height=1.25in,clip,keepaspectratio]{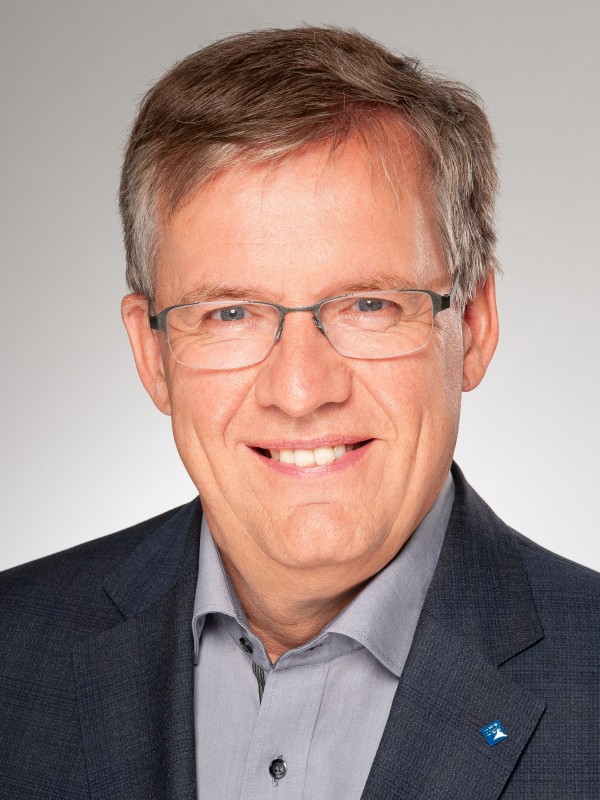}}]{Jörn Ostermann} (Fellow, IEEE) received the degree in electrical engineering from the University of Hannover, the degree in communications engineering from Imperial College London, and the Dr.-Ing. degree from the University of Hannover, in 1994, for his work on low bit-rate and object-based analysissynthesis video coding. Since 2003, he has been a Full Professor and the Head of the Institut für  Informationsverarbeitung, Leibniz Universität Hannover, Germany. In July 2020, he was appointed as a Convenor of the MPEG Technical Coordination. He is named as an inventor on more than 30 patents. His current research interests include video coding and streaming, computer vision, machine learning, 3D modeling, face animation, and computer-human interfaces. He is a member of the IEEE Technical Committee on Multimedia Signal Processing and the past Chair of the IEEE CAS Visual Signal Processing and Communications (VSPC) Technical Committee.
\end{IEEEbiography}

\vfill

\end{document}